\documentclass[useAMS,usenatbib,usegraphicx]{mn2e}

\title[Extremely red galaxies: dust attenuation and classification]{Extremely red galaxies: dust attenuation and classification}
\author[D. Pierini C. Maraston R. Bender and A. N. Witt]{D. Pierini$^{1}$\thanks{E-mail: dpierini@mpe.mpg.de}, C. Maraston$^{1}$, R. Bender$^{1,2}$
and A. N. Witt$^{3}$\\
$^{1}$Max-Planck-Institut f\"ur extraterrestrische Physik, Giessenbachstr., D-85748 Garching, Germany\\
$^{2}$Universit\"ats-Sternwarte M\"unchen, Scheinerstr. 1, D-81679 M\"unchen, Germany\\
$^{3}$Ritter Astrophysical Research Center, The University of Toledo, Toledo, OH 43606, U.S.A.}
\begin{document}

\date{Accepted ... . Received ... ; in original form ...}

\pagerange{\pageref{firstpage}--\pageref{lastpage}} \pubyear{2002}

\maketitle

\label{firstpage}

\begin{abstract}
We re-address the classification criterion for extremely red galaxies (ERGs)
of Pozzetti \& Mannucci (2000 -- PM00), which aims to separate,
in the $\rm I_c-K$ (or $\rm R_c-K$) vs. $\rm J-K$ colour--colour diagram,
passively evolving, old ($\rm \ge 1~Gyr$) stellar populations
in a dust-free environment, associated with ellipticals (Es),
from dusty starburst galaxies (DSGs), both at $1 < z < 2$.
We explore a category of objects not considered previously,
i.e., galaxies forming {\it in this redshift range} on short
($\rm 0.1~Gyr$) timescales and observed also in their early,
dusty post-starburst phase.
We also investigate the impact of structure of the dusty medium
and dust amount on the observed optical/near-IR colours of high-$z$ DSGs/DPSGs,
through multiple-scattering radiative transfer calculations
for a dust/stars configuration and an extinction function calibrated
with nearby dusty starbursts.
As a main result, we find that dusty post-starburst galaxies (DPSGs),
with ages between 0.2 and 1 Gyr, at $1.3 < z < 2$ mix with Es at $1 < z < 2$
for a large range in dust amount.
This ``intrusion'' is a source of concern for the present two-colour
classification of ERGs.
On the other hand, we confirm, in agreement with PM00,
that DSGs are well separated from Es, both at $1 < z < 2$,
in the $\rm I_c-K$ vs. $\rm J-K$ colour--colour diagram,
whatever the structure (two-phase clumpy or homogeneous) of their dusty medium
and their dust amount are.
This result holds under the new hypothesis of high-$z$ Es being as dusty
as nearby ones.
Thus the interpretation of the optical/near-IR colours of high-$z$ Es
may suffer from a multiple degeneracy among age, metallicity, dust
and redshift.
We also find that DPSGs at $z \sim 1$ mix with DSGs at $1 < z < 2$,
as a function of dust amount and structure of the dusty medium.
All these results help explaining the complexity of the ERG classification
emerged from recent surveys.
Model DSGs/DPSGs with $\rm K < 20.5$ and selected as ERGs on the basis of
their observed optical/near-IR colours span a large range
in absorption efficiency and/or intrinsic UV-to-optical luminosity ratio.
As a consequence, the total emission from dust in the rest-frame
mid-IR--(sub)mm ranges from $\sim 10^{10}$
to $\rm \sim 5.6 \times 10^{12}~L_{\sun}$ for DPSGs through DSGs.
This large dynamic range is consistent with estimates
made from submm observations of bright ($\rm K < 19.5$--20) ERGs.
Finally, we show that ERG samples are biased towards a lower fraction
either of DSGs/DPSGs at $1.3 < z < 2$ or of DSGs/DPSGs at $z \sim 1$,
when selection is made via either the $\rm R_c-K \ge 5.3$
or the $\rm I_c-K \ge 4$ colour criterion.
\end{abstract}


\section{Introduction}

The extremely red objects (EROs), originally discovered in the K-band survey
of Elston, Rieke \& Rieke (1988), have become more and more present
in deep near-IR and optical surveys (see Yan \& Thompson 2003 for a review).
Operationally defined in terms of the observed optical/near-IR colours
(e.g. $\rm R_c-K \ge 5.3$ or $\rm I_c-K \ge 4$), significantly different
from those of typical field objects, EROs may nevertheless comprise
an assorted mix of Galactic and extragalactic objects.
We focus on the extremely red galaxies (ERGs), which account for the bulk
of the observed EROs.
For the optically selected ERGs, the contamination from AGNs is less than
15 per cent (Alexander et al. 2002; Brusa et al. 2002; Smail et al. 2002),
even though obscured, Compton-thin AGNs, partly associated with
dusty starbursts having large star formation rates (SFRs),
most probably dominate X-ray selected ERG samples (Stevens et al. 2003).

ERGs still lack an uncontroversial classification because
their very red colours can be reproduced by at least two very different sets
of spectral energy distributions (SEDs) not associated with an AGN:
those expected for passively evolving, old ($\rm \ge 1~Gyr$)
stellar populations in a dust-free environment, associated with ellipticals,
and those expected for dusty starburst galaxies (DSGs),
both at high redshift ($z > 1$).
Indeed it emerges that a half or more of the entire ERG sample
in the literature contains ellipticals (Moriondo, Cimatti \& Daddi 2000;
Stiavelli \& Treu 2001; Mannucci et al. 2002; Cimatti et al. 2002a).

ERGs represent a key-case in the knowledge of the cosmic history
of star formation.
In fact old, massive ellipticals at $z > 1$ represent a problem for
the standard hierarchical merging scenario of galaxy formation and evolution
(White \& Frenk 1991), where the most massive galaxies are assembled through
a merging tree of less massive systems, even though the bulk of their stars
may already be very old at the time of the last major merging
(e.g. Kauffmann, Charlot \& White 1996; Springel \& Hernquist 2003).
Conversely, the association of ERGs with DSGs at intermediate redshifts
does not necessarily favour the latter scenario, if galaxies are assembled
within times comparable to their star formation timescales
and the latter are shorter the larger the masses of the galaxies are,
as independently proposed for late- (Gavazzi, Pierini \& Boselli 1996)
and early-type galaxies (Thomas, Maraston \& Bender 2002).
Interestingly ``hyper'' ERGs have been interpreted as primodial ellipticals
in a dusty starburst phase at $z \sim 3$ (Totani et al. 2001).

Unfortunately the classification either as a passively evolving, old elliptical
or as a dusty starburst galaxy at $z > 1$ is uncomfortably ambiguous
in absence of an extensively observed SED (e.g., Smith et al. 2001),
but the needed observations are extremely challenging at present,
given the faintness of the ERGs (e.g. Cimatti et al. 1999).
Hence Pozzetti \& Mannucci (2000 -- hereafter referred to as PM00)
have proposed a classification criterion based on the relatively easily
obtainable optical/near-IR photometry, particularly suitable for large samples
of EROs with $\rm K < 20$--21.
This classification criterion is based on the rather neat difference
in the distribution of the PM00 model ellipticals and DSGs at $1 < z < 2$
in the $\rm I_c-K$ (or $\rm R_c-K$) vs. $\rm J-K$ colour--colour diagram.
This separation is mainly due to the fact that passively evolving, old
stellar populations at these redshifts have their strong $\rm 4000~\AA$ break
redshifted to $\rm \lambda < 1.2~\mu m$ (J band),
while coeval dusty galaxies show smoother SEDs and therefore redder
observed $\rm J-K$ colours, whatever the dust/stars configuration is.

For the dusty ERGs, there is no unanimous consensus on the typical dustiness
and star formation rate.
The detection of ERO\,J164502+4626.4 ($z=1.44$) at $\rm 850~\mu m$
(Cimatti et al. 1998) suggested that dusty ERGs may host massive quantities
of dust accompanied by very high SFRs, as in ultraluminous infrared galaxies
(ULIRGs) (see also Cimatti et al. 1997; Elbaz et al. 2002).
Independently, 1.4 GHz radio maps reveal that $\ge 16 \pm 5$ per cent
of the ERGs with $\rm K < 20.5$ have radio luminosities expected for ULIRGs
at $z \sim 1$ (Smail et al. 2002).
Conversely, Andreani et al. (1999) and Mohan et al. (2002) conclude that
about 20 per cent of the bright ($\rm K < 19.5$--20) ERGs show
strong submm emission but that the dominant population of dusty, bright ERGs
is probably made of starburst galaxies with a total mid-IR--mm
rest-frame luminosity lower than $10^{12}~\rm L_{\odot}$
and $\rm SFR <$ few $\rm \times 10^2~M_{\odot}~yr^{-1}$ (i.e., non-ULIRGs).

Here we re-address the optical/near-IR photometric classification of ERGs
and investigate the relation between reddening, dustiness
and SFR in dusty ERGs.
Four are the main new points of this study.
First, we consider the presence of {\it stellar populations
which are formed in a burst of short ($\rm 0.1~Gyr$) duration
and are observed at $1 < z < 2$ in an early post-starburst phase
while still embedded in a dusty medium}.
Second, we consider the presence of passively evolving, old,
{\it dusty elliptical galaxies} at $1 < z < 2$ as additional candidate ERGs.
Third, the treatment of dust attenuation follows the Monte Carlo simulations
of the radiative transfer by Witt \& Gordon (2000 -- hereafter referred to
as WG00) for starbursts and post-starbursts, and those of
Witt, Thronson \& Capuano (1992 -- hereafter referred to as WTC92)
for ellipticals.
{\it For DSGs and dusty post-starburst galaxies (DPSGs), these models explore
the dependence of dust attenuation from wavelength as a function of
a dust/stars configuration, structure of the dusty medium and dust properties
which were not considered simultaneously or at all by PM00} (see Sect. 2.2.2).
Fourth, we calculate the total emission from dust in the mid-IR--(sub)mm
wavelength range for DSGs/DPSGs.

\section{Modelling passively evolving, old elliptical galaxies
and dusty starburst/post-starburst galaxies at $\bf 1 < z < 2$}

For our candidate ERGs (passively evolving, old ellipticals
and DSGs/DPSGs), we consider three representative redshifts
($z = 1.07$, 1.39 and 1.85), consistent with published results.
We adopt the so-called ``concordance'' cosmological model
($\Omega_m = 0.3$, $\Omega_{\lambda} = 0.7$) with Hubble constant
$H_0 = 70~\rm km~s^{-1}~Mpc^{-1}$ to link redshift and age of the models.

We use two evolutionary synthesis codes
(one for ellipticals and one for DSGs/DPSGs)
and fix metallicity (i.e., solar -- $\rm Z_{\sun}$)
and initial mass function (Salpeter 1955),
since PM00 made an exhaustive investigation of the influence of
these parameters on the colours of their candidate ERGs.
However we will discuss the impact of sub-solar metallicities on our results.

Our focus is on the $\rm I_c-K$, $\rm R_c-K$ and $\rm J-K$ colours,
where Cousins (1978) $\rm I_c$ and $\rm R_c$ filters
and Bessel \& Brett (1988) J and K filters are used.

\subsection{Passively evolving, old elliptical galaxies}

\subsubsection{Stellar populations}

As in PM00, passively evolving, old ($\rm \ge 1~Gyr$) elliptical galaxies
are simulated through simple stellar population (SSP) models.
Hereafter they are referred to as Es for simplicity.
Their intrinsic (i.e., unattenuated by dust) SEDs are computed with
the evolutionary synthesis code of Maraston (1998).

We assume a maximal formation redshift $z_f = 10$, against the three values
of formation redshift (i.e., $z_f = 3$, 4 and 6) assumed by PM00
for their model Es.
An elliptical galaxy formed at $z_f = 10$ has an age of about 5, 4 and 3 Gyr
at $z = 1.07$, 1.39 and 1.85, respectively.
One formed at $z_f = 3$ (PM00; see also Totani et al. 2001) has an age
of about 3.4, 2.4 and 1.4 Gyr at $z = 1.07$, 1.39 and 1.85, respectively.

We note that a 1 Gyr-old galaxy at $z = 1.07$, 1.39 or 1.85
was formed at $z_f = 1.39$, 1.85 or 2.59, respectively.
Thus a 1 Gyr-old elliptical at $1 < z \le 1.5$ may have its progenitor
in the redshift range that we are interested in.
Hence it is important to investigate the case of a post-starburst galaxy
and illustrate the difference between a model DPSG
and a model elliptical galaxy of similar young/intermediate age
(see Sect. 2.2.1).

For model Es we adopt a minimum stellar mass of
$\rm 7.5 \times 10^{10}~M_{\sun}$, which is about the residual
(i.e., after 25 per cent mass-loss owing to stellar evolution) stellar mass of
a 1 Gyr-old SSP (from Maraston 1998) as well as of a 1 Gyr-old model DSG/DPSG
(cf. Sect. 2.2.1), both originating from a $\rm 10^{11}~M_{\sun}$ cloud of
primordial gas.

\subsubsection{Dust attenuation}

We test the impact of dust attenuation on the observed SEDs
of Es at $1 < z < 2$.
The residual interstellar medium of a dusty elliptical is considered
to be negligible with respect to its characteristic stellar mass.

For dusty Es we adopt the dust attenuation curves obtained from
the Monte Carlo simulations of the radiative transfer by WTC92
for the CLOUDY geometry (see also WG00).
This geometry assumes a spherical stellar distribution that declines from
the core as $r^{-3}$ and a {\it homogeneous Milky Way-type dust present only
in the inner portions of the model}, extending from the center to two thirds
of the galaxy radius and following a distribution that declines as $r^{-1}$.
The results of this model are robust when different radial dependences
of the stellar and dust distributions are assumed
(cf. Bianchi, Ferrara \& Giovanardi 1996; Wise \& Silva 1996; WG00).

The value of the radial extinction optical depth from the center
to the edge of the dust environment at V-band, assuming a constant density,
homogeneous distribution, ($\tau_V$) is set to 0.5, consistent with
the mean $\tau_V$ estimated from the analysis of broad-band colour gradients
in nearby Es (WTC92; Wise \& Silva 1996).
These studies provide upper limits to $\tau_V$, since they do not take
into account intrinsic gradients in metallicity of the stellar populations
of Es (e.g. Saglia et al. 2000).

In Fig. 2, we reproduce the attenuation function\footnote{We stress
the difference between dust extinction curve and dust attenuation curve.
The former describes the absorption and scattering properties of a mix of
dust grains of given size distribution and chemical composition
as a function of wavelength; the latter is the convolution of
the extinction curve with the geometry of the dusty stellar system.
The attenuation curve depends on structure of the dusty medium
and dust amount.}, normalized to its V-band value, (${A_{\lambda}}^{norm}$)
of dusty Es for $\tau_V = 0.5$, which produces a value of the stellar reddening
($E(B-V)_{star}$) equal to 0.15.

\subsection{Dusty starburst/post-starburst galaxies}

\subsubsection{Stellar populations}

Intrinsic SEDs of young ($\rm \le 0.1~Gyr$)/intermediate-age (0.1--1 Gyr)
starburst and intermediate-age post-starburst galaxies are computed with
the evolutionary synthesis code (version 2.0) P\'EGASE
(Fioc \& Rocca-Volmerange 1997), because this code includes nebular emission.
P\'EGASE 2.0 does not include the TP-AGB phase that is dominant
at intermediate ages, but this fact does not represent a major problem
since we are considering observed wavelengths shorter than $\rm 2.2~\mu m$
(K band).

Gas is assumed to be transformed into $\rm Z_{\sun}$ stars
at a constant rate over a period $t_{burst}$ of $\rm 0.1~Gyr$
({\it short burst}) or $\rm 1~Gyr$ ({\it long burst}) at $1 < z < 2$.
For a total mass of transformed gas $M_{gas} = \rm 10^{11}~M_{\sun}$,
a SFR equal to $M_{gas} / t_{burst}$ corresponds to $\rm 10^2~M_{\sun}~yr^{-1}$
(long burst) or $\rm 10^3~M_{\sun}~yr^{-1}$ (short burst).
Note that SFRs $\rm \sim 10^3~M_{\sun}~yr^{-1}$ have been inferred
from observations of high-$z$ SCUBA submm sources
(e.g., Smail, Ivison \& Blain 1997; Hughes et al. 1998; Barger et al. 1998;
Cimatti et al. 1998; Eales et al. 1999; Bertoldi et al. 2000; Lutz et al. 2001;
Smail et al. 2003).

These star formation histories and SFRs are complementary
to those adopted by PM00 for their DSGs for the following reason.
{\it The constant star formation rate from a formation redshift 2.5
or, especially, 6, assumed by PM00 for their basic model of a DSG,
better corresponds to a continuously star-forming galaxy
with an intermediate/low SFR at $1 < z < 2$}.
An object formed at $z_f =2.5$(6) is about 0.7(2.3) Gyr old at $z=2$,
and 3.2(4.8) Gyr old at $z=1$, for the same cosmology as in Sect. 2.
Hence it may hardly sustain a constant SFR larger than
$\rm 10^2~M_{\sun}~yr^{-1}$ for such long times.
For our DSGs/DPSGs observed at $z = 1.07$, 1.39 or 1.85, the maximum age
of 1 Gyr corresponds to a maximum $z_f$ of 1.39, 1.85 or 2.59, respectively.
Hence {\it these systems are mostly formed at $1 < z < 2$}.
We consider additional model ages of 50, 100, 200, 300 and 500 Myr.

PM00 consider also model starbursts with a SFR having an e-folding time
of 100 Myr.
These PM00 models are consistent with our models of starbursts
with a short burst at an age of 100 Myr.
However, PM00 do not consider an early post-starburst phase at $1 < z < 2$.

At this point the difference between a post-starburst galaxy
and a passively evolving elliptical galaxy younger than 1 Gyr
may sound a matter of semantics, since a DSG at $1 < z < 2$ may evolve
into an elliptical at $z=0$ and, eventually, already at these redshifts.
As far as stellar properties are concerned, the SEDs of
the post-starburst stellar population under investigation
and of an SSP with the same age and metallicity, identified as
a passively evolving elliptical galaxy, are different (as expected)
even after 1 Gyr from the onset of star formation, as shown in Fig. 1.
However, it is the presence, distribution and structure of
the (dusty) interstellar medium which provides fundamental physical differences
between these two stellar systems in our models.

The mass fractions of stars and residual gas are 74 (77) per cent
and 22 (19) per cent, respectively, for a 1 Gyr-old stellar system
formed in a short (long) burst.
For DSGs with a short burst, these fractions are 87 per cent and 11 per cent,
respectively, at the end of the burst (i.e., at an age of 100 Myr).
We assume that gas and dust components survive at least partly
in model DSGs/DPSGs for the largest time-window (1 Gyr) considered
for these objects (see Sect. 2.2.3).
The fate of these components afterwards may channel the evolution
of these starbursts towards either a ``gas-poor'' system
(i.e., an early-type galaxy) or a ``gas-rich'' one (i.e., a late-type galaxy)
at $z=0$, both systems having a stellar mass of
$\rm \sim 10^{10}$--$\rm 10^{11}~M_{\sun}$.

Thus a 1 Gyr-old DPSG at $z = 1.85$ could be the transition evolutionary phase
of elliptical dusty formation occurred at intermediate redshift ($z = 2$--3)
(Totani et al. 2001; Granato et al. 2001).
However the dusty formation of more massive Es at intermediate redshift
would require initial starbursts with $\rm SFR \ge 10^3~M_{\sun}~yr^{-1}$
and/or $t_{burst} \ge \rm 100~Myr$ (cf. Granato et al. 2001).
\begin{figure}
  \includegraphics[width=84mm]{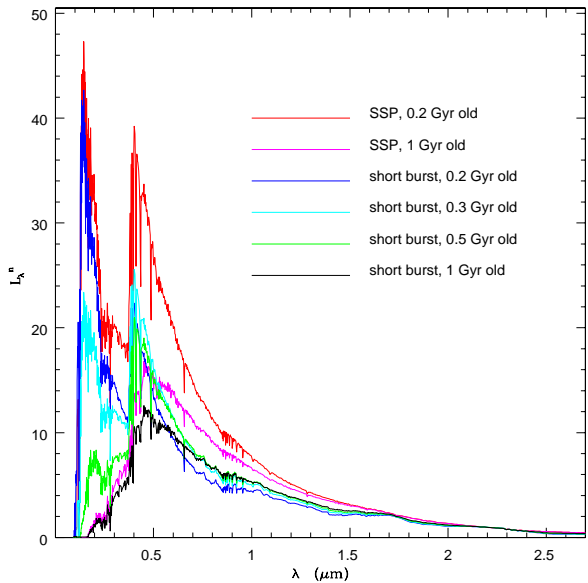}
  \caption{The intrinsic SED of an SSP with $\rm Z_{\sun}$ and age equal to
0.2 (red line) or 1 Gyr (magenta line) and the intrinsic SED
of a stellar population originated in a 100 Myr burst with $\rm Z_{\sun}$
and age equal to 0.2 (blue line), 0.3 (cyan line), 0.5 (green line)
or 1 Gyr (black line). Each SED is in the object's rest-frame and is normalized
to the luminosity at $\rm 2.17~\mu m$. It is evident that these two sets
of SEDs are different at any intermediate age.}
  \label{Fig. 1}
\end{figure}

\subsubsection{Dust attenuation}

Gordon, Calzetti \& Witt (1997 -- hereafter referred to as GCW97)
have investigated the nature of the dust in starbursts
by constructing different models of the stars and dust in DSGs,
combining stellar evolutionary synthesis and radiative transfer through dust,
and applying them to the observed UV/optical colours of
30 nearby starburst galaxies of the Calzetti sample
(Calzetti, Kinney \& Storchi-Bergmann 1994).
Their main conclusion are as follows.
\begin{itemize}
\item
Starburst dust has an extinction curve lacking the $\rm 2175~\AA$ bump,
like the Small Magellanic Cloud (SMC) curve and at variance with
the Milky Way (MW) curve, and a steep far-UV rise, intermediate between
these two curves.
This is consistent with the conclusion of Calzetti et al..
\item
The dust/stars geometry that better explains the distribution
of these 30 starbursts in various UV/optical colour--colour plots
has an inner dust-free sphere of stars surrounded by an outer star-free shell
with a two-phase clumpy, dusty medium.
A geometry where dust and stars are uniformly mixed through the entire sphere
can not explain the spread in the UV/optical colours,
be its local distribution of dust homogeneous or clumpy.
\end{itemize}
Models of dusty starbursts based on similar considerations
well reproduce the optical colour--redshift distribution of
Lyman-Break Galaxies at $2 < z < 4$ (Vijh, Gordon \& Witt 2003).
In addition, Cimatti et al. (1997) suggest that the stellar component
of ERO\,J164502+4626.4 should be strongly embedded in the dust
in order to reproduce its observed extremely red colour.

For all these reasons, we apply the Monte Carlo simulations of
the radiative transfer by WG00 for the SHELL geometry to DSGs/DPSGs
at $1 < z < 2$.
This geometry assumes {\it a dust/stars distribution where the stars extend to
0.3 of the system radius and the dust extends from 0.3 to 1
of the system radius}.
The dust in the shell is distributed in a {\it two-phase clumpy medium},
comprised of molecular clouds and diffuse gas,
where the filling factor of the clumps is set to 0.15
and the low-density to high-density ratio is set to 0.01.
We consider the additional extreme hypothesis that dust
is locally distributed in a single, {\it homogeneous medium}.

We assume an {\it SMC-like extinction curve} and consider values of $\tau_V$
equal to 0.25, 0.5, 1, 2, 3, 4, 5, 6, 8, 10, 20, 30, 40 and 50
for the two-phase clumpy dust distribution and equal to 0.25 through 8
for the homogeneous one.
Here we recall that {\it $\tau_V$ gives the total amount of dust} in the shell,
since the dust column density is proportional to $\tau_V \times a \times \rho$,
where $a$ is the characteristic grain size and $\rho$ is the density
of the grain material.
Values of $\tau_V$ at the low- and high-end of the distribution
seem to be characteristic of starbursts and ULIRGs, respectively,
at low- as well as high-$z$ (e.g. Vijh et al. 2003 and references therein).

We also assume that the gas emission at a given wavelength is attenuated
by the same amount as the stellar emission at that wavelength,
whether the gas emission is either in a line or in the continuum.
This is at odds with the result that the stellar continuum suffers roughly
half of the reddening suffered by the ionized gas (Calzetti et al. 1994;
Fanelli, O'Connell \& Thuan 1988; Mas-Hesse, Arnault \& Kunth 1989)
but the understanding of the attenuation of the ionized-gas line emission
from dust is still controversial.
However, {\it the following results hold when no nebular emission
is taken into account}, as demonstrated in App. A.

Fig. 2 reproduces the attenuation curves (normalized to V band)
for the SHELL dust/stars configuration and SMC-type dust locally distributed 
in a homogeneous (solid lines) or two-phase clumpy medium (dotted lines)
as a function of $\tau_V$. 
It also shows the values of the normalized attenuation function
of nearby starbursts (Calzetti et al. 2000), known as ``Calzetti law'',
for the WG00 wavelengths (filled triangles).

Two important results are contained in Fig. 2.
First, the attenuation function for the SHELL geometry with a two-phase clumpy
distribution of SMC-type dust and $\tau_V = 1.5$ reproduces rather well
the calibrated Calzetti law for nearby starbursts, as already shown
by Vijh et al. (2003).
Nevertheless a wide range of different attenuation curves may apply
to individual DSGs/DPSGs, as a function of local distribution of the dust
and dust content.
We discuss the effect of using the empirically determined Calzetti law
for whatever dusty ERG (e.g. Cimatti et al. 2002a) in App. B.

Second, {\it the shape of the attenuation function depends on
structure of the dusty medium and, in the two-phase clumpy case, on $\tau_V$}.
Changes in normalization also exist, in both cases, of course (not shown).
In particular, a homogeneous dusty medium provides a larger attenuation
at any wavelength than a two-phase clumpy one.

In addition, we stress that {\it the relation between $E(B-V)_{star}$
and dust amount (i.e., $\tau_V$) is usually quite non linear
and highly dependent on details of the structure of the dusty medium},
as shown in Fig. 3.
The existence of an initial ``plateau'' in $E(B-V)_{star}$ vs. $\tau_V$
in the two-phase clumpy case is due to fact that individual clumps
become optically thick while the low-density interclump medium
initially contributes little to the reddening.

Finally, we note that, apart from the observationally-established
attenuation function of Calzetti (1997) (see App. B),
all the model attenuation functions considered by PM00 do not contemplate
dust with SMC-type extinction curve and distributed in a two-phase
clumpy medium {\it at the same time}, as well as a SHELL-like geometry.
We anticipate that these important differences do not affect our re-analisys of
the two-colour classification criterion for ERGs.
In fact our results are consistent with those of PM00
when continuously star-forming DSGs are considered, whatever structure
of the dusty medium is assumed.

\subsubsection{The physics behind the SHELL configuration}
\begin{figure}
  \includegraphics[width=84mm]{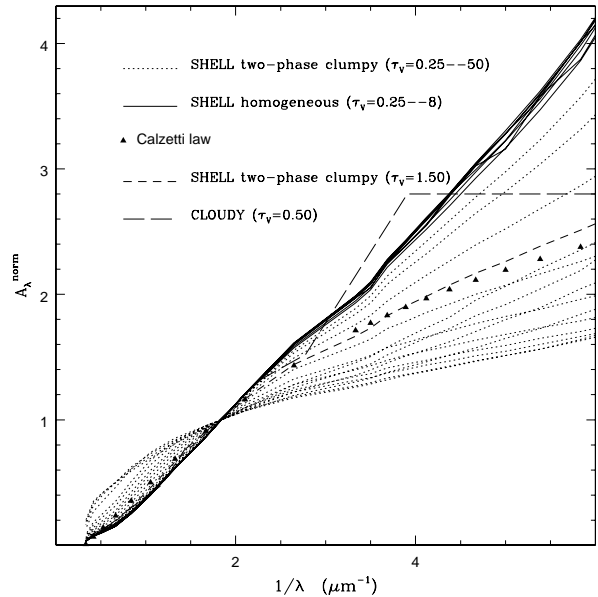}
  \caption{The attenuation function, normalized to its V-band value,
(i.e., ${A_{\lambda}}^{norm}$) for: (a) the WTC92 CLOUDY geometry
of a dusty elliptical galaxy with MW-type dust and $\tau_V = 0.5$
(long-dashed line); (b) the WG00 SHELL geometry of a DSG with SMC-type dust
(solid or dotted lines for a homogeneous or two-phase clumpy dust distribution,
respectively) and different values of $\tau_V$. As a reference, we reproduce
the normalized attenuation function for the WG00 SHELL model
with clumpy SMC-type dust and $\tau_V = 1.5$ (short-dashed line),
which is consistent with the normalized Calzetti law (filled triangles).}
  \label{Fig. 2}
\end{figure}
\begin{figure}
  \includegraphics[width=84mm]{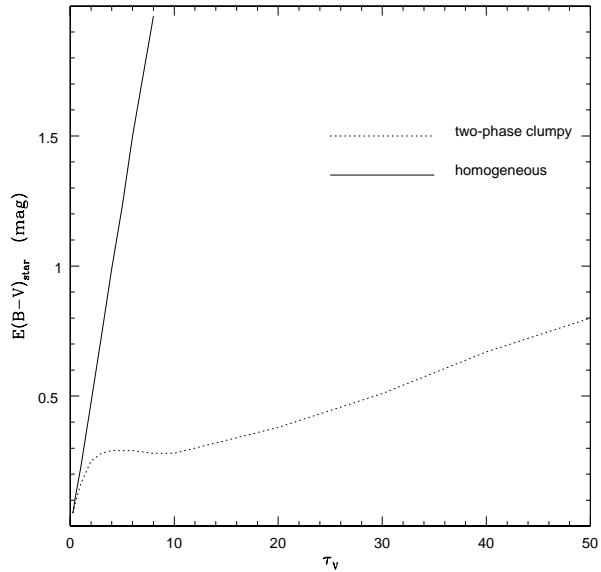}
  \caption{Stellar reddening $E(B-V)_{star}$ vs. $\tau_V$
for the SHELL geometry of a DSG with SMC-type dust distributed in a homogeneous
(solid line) or two-phase clumpy (dotted line) medium.}
  \label{Fig. 3}
\end{figure}

The existence of dust in young galaxies is expected because Type II supernovae
(SNe II) are shown to produce dust grains (e.g., Dwek et al. 1983;
Moseley et al. 1989; Kozasa, Hasegawa \& Nomoto 1991; Todini \& Ferrara 2001).
Dust production from SNe II follows the formation and evolution
of massive stars on a nuclear timescale (few $\rm \times 10^{6}~yr$).
Mass loss from evolved intermediate-mass stars contributes to dust formation
as well (Gehrz 1989) and follows the slower evolutionary timescales
of these stars (from $\rm \sim 10^{8}$ to $\rm \sim 10^{9}~yr$).
Dust is also destroyed by SN shocks (McKee 1989;
Jones, Tielens \& Hollenbach 1996).
Nevertheless, the detailed modelling of dust evolution in high-$z$ and nearby
young starbursts shows that it is possible to build up a dust mass
within a factor of two from total within a few $\rm \times 10^{7}~yr$
(Hirashita \& Ferrara 2002; Hirashita, Hunt \& Ferrara 2002).
Early dust is similar to the SMC one (Takeuchi et al. 2003).

Dust and residual gas face displacement and, eventually,
partial/total removal from the system owing to the onset of galactic winds
powered by SN explosions and accelerated up to velocities of
several $\rm \times 10^3~km~s^{-1}$ (e.g. Suchkov et al. 1994;
Suchkov et al. 1996; Shu, Mo \& Mao 2003).
In the case of dust grains, destruction by sputtering must be taken
into account (e.g. Ellison, Drury \& Meyer 1997).
SN Ia will contribute to dust destruction/displacement as well as SN II,
because their timescales span a wide range: roughly 60 per cent of
Type Ia SN progenitors have lifetimes $\sim 3$--$\rm 4 \times 10^8~yr$,
while 30 per cent of them are expected to be very old
($10^9$--$\rm 10^{10}~yr$) (Branch et al. 1995).
The existence of galactic superwinds associated with dust extends
from nearby to high-$z$ starbursts (Heckman 2001 and references therein;
Adelberger et al. 2003).

The fate of the surviving ejected dust grains is uncertain
for the following reasons.
The ambient halo gas that the superwind shocks and the dense disk-gas
that it entrains travel at velocities of a few $\rm \times 10^2~km~s^{-1}$,
which are near to typical escape velocities (Nakai et al. 1987; Phillips 1993).
Furthermore, displaced grains may respond as much to radiation pressure
(Davies et al. 1998; Norman \& Ferrara 1996) as they do to gas pressure
(as well as to any poloidal magnetic field that might be present --
Beck et al. 1994).
Hence displaced dust may decouple from the gas so that its path through
the halo becomes even more uncertain.
Displaced grains may partially undergo both reconnection to the original disk
and removal from the system, as well as they may exist in equilibrium positions
above the plane of the galaxy (Davies et al. 1998; see also Greenberg et al.
1987 and Barsella et al. 1989).
This supports our assumption that the SHELL geometry applies also to
the immediate post-burst phase of a DSG.

\section{Prerequisites for a model being an ERO}
\begin{figure}
  \includegraphics[width=84mm]{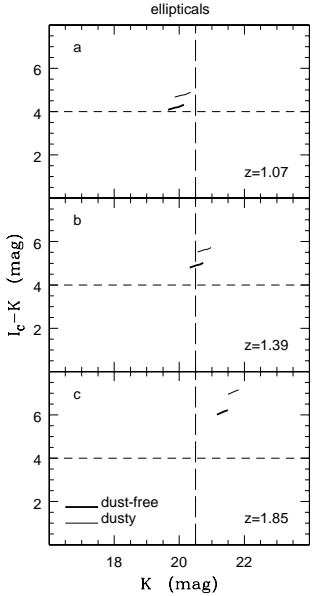}
  \caption{Distribution of model Es at $z = 1.07$ ({\bf a}), $z = 1.39$
({\bf b}) or $z = 1.85$ ({\bf c}) in the $\rm I_c-K$ vs. K colour--magnitude
diagram for the dust-free (thick solid line) or dusty (thin solid line) case.
In each panel, the model age increases from left to right along each line.
Long- and short-dashed lines reproduce the limiting magnitude of current
ERO samples and the $\rm I_c-K$ colour selection criterion for EROs,
respectively. Note the extremely narrow locus populated by Es at $1 < z < 2$
in this plane.}
  \label{Fig. 4}
\end{figure}
\begin{figure}
  \includegraphics[width=84mm]{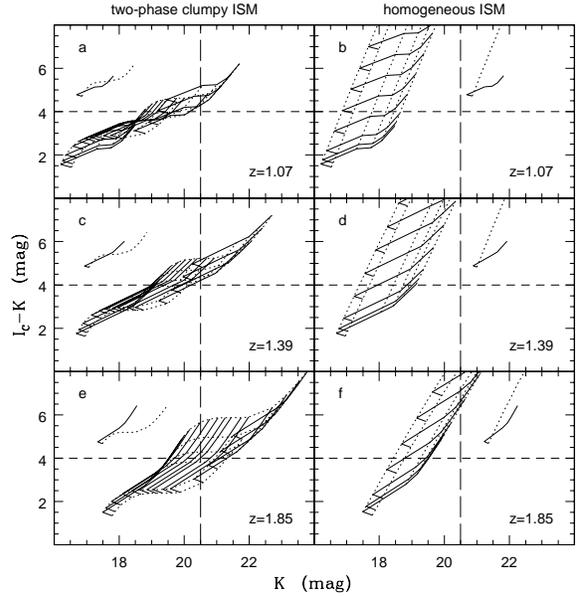}
  \caption{Distribution of model DSGs with a short burst and DPSGs
at $z = 1.07$ ({\bf a}, {\bf b}), $z = 1.39$ ({\bf c}, {\bf d}) or $z = 1.85$
({\bf e}, {\bf f}) in the $\rm I_c-K$ vs. K colour--magnitude plane
for the two-phase clumpy ({\bf a}, {\bf c}, {\bf e}) and homogeneous
({\bf b}, {\bf d}, {\bf f}) dust distributions. Long- and short-dashed lines
are the same as in Fig. 4. Solid or dotted lines connect models with
constant $\tau_V$ or age, respectively. In the upper left (right) corners of
panels a,c,e (b,d,f) we reproduce the typical trend of models with
increasing age (solid line) or increasing $\tau_V$ (dotted line) -- see text.}
  \label{Fig. 5}
\end{figure}
\begin{figure}
  \includegraphics[width=84mm]{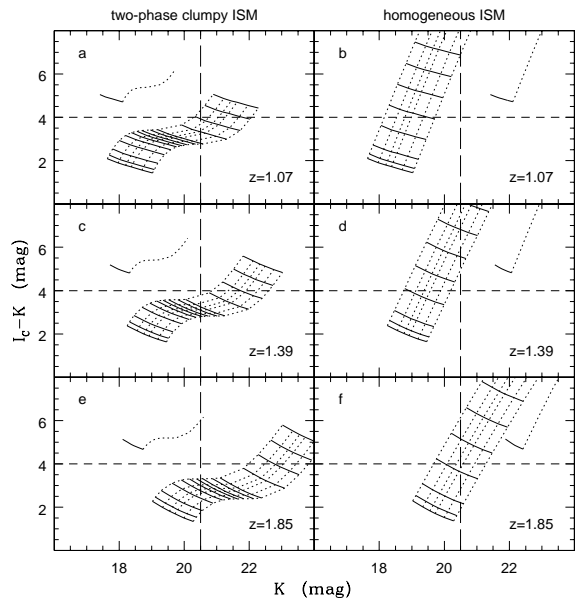}
  \caption{The same as in Fig. 5 but for model DSGs with a long burst.
Solid or dotted lines connect models with constant $\tau_V$ or age,
respectively. In the upper left (right) corners of panels a,c,e (b,d,f)
we reproduce the typical trend of models with increasing age (solid line)
or increasing $\tau_V$ (dotted line) -- see text.}
  \label{Fig. 6}
\end{figure}

Existing ground-based near-IR surveys (Drory et al. 2001; Cimatti et al. 2002b;
Smail et al. 2002) reach a typical limit of $\rm K = 20.5$.
$\rm K < 20$--21 is also the limit of validity of the PM00 method,
introduced by these authors in order to avoid statistical contamination
by $z > 2$ ellipticals in the optical/near-IR colour--colour region
populated by DSGs.

An important prerequisite of our models is to match
observed photometric properties of EROs, i.e. $\rm K < 20.5$
and e.g. $\rm 4 \le I_c-K < 8$.
Thus we show the distribution in the $\rm I_c-K$ vs. K
colour--magnitude plane of our models of Es, DSGs with a short burst
and DPSGs, and DSGs with a long burst in Figs. 4, 5 and 6, respectively.

Here we remind that our model Es have a minimum stellar mass equal to
$\rm 7.5 \times 10^{10}~M_{\sun}$ whatever their age (from 1 to 3--5 Gyr) is.
Model DSGs with a short burst have stellar masses equal to 4.5
and $\rm 8.7 \times 10^{10}~M_{\sun}$ when they are 0.05 and 0.1 Gyr old,
respectively, while model DPSGs have stellar masses equal to 8.1, 7.9, 7.7
and $\rm 7.4 \times 10^{10}~M_{\sun}$ when they are 0.2, 0.3, 0.5
and 1 Gyr old, respectively.
Finally, DSGs with a long burst have stellar masses equal to 0.4, 0.9,
1.7, 2.5, 4.0 and $\rm 7.7 \times 10^{10}~M_{\sun}$
when they are 0.05, 0.1, 0.2, 0.3, 0.5 and 1 Gyr old, respectively.

From Fig. 4 it is evident that model dust-free Es
of $\rm 7.5 \times 10^{10}~M_{\sun}$ correspond to observed EROs
only if they are at $z < 1.4$ (for any age at $z \sim 1$
or for ages less than 2 Gyr at $z \sim 1.4$).
Dusty Es of the same stellar mass correspond to observed EROs
only if they are at $z \sim 1$ (for any age).
Hence stellar mass plays a key role: a ten times more massive
passively evolving, old elliptical galaxy at $1< z < 2$
is always a candidate ERO, whether dusty or not.
Indeed large near-IR surveys aimed at testing galaxy formation models
are designed to detect Es more massive than $\rm \sim 10^{11}~M_{\sun}$
at high-$z$ (e.g. Drory et al. 2001).

The prerequisites for a model DSG/DPSG at $1< z < 2$ being a candidate
observed ERO are more complex, as illustrated in Figs. 5 and 6
and summarised in Table 1.
This is largely due to the well-known existence of the age/opacity degeneracy
(Takagi, Arimoto \& Vansevicius 1999 and references therein).
\begin{table}
 \centering
  \caption{Parameters of model DSGs/DPSGs selected as EROs with $\rm K < 20.5$}
  \begin{tabular}{@{}rrccc@{}}
  \hline
   $\rm t_{burst}$ & age & two-phase clumpy & homogeneous & z \\
   Myr & Myr & & & \\
  \hline
   100 & 50 & $\rm \tau_V > 40$ & $\rm 3 < \tau_V < 8$ & 1.07 \\
   100 & 100 & $\rm \tau_V \ge 40$ & $\rm 3 \le \tau_V < 8$ & 1.07 \\
   100 & 200 & $\rm  20 < \tau_V < 50$ & $\rm 2 < \tau_V < 8$ & 1.07 \\
   100 & 300 & $\rm  20 < \tau_V < 40$ & $\rm 2 < \tau_V < 8$ & 1.07 \\
   100 & 500 & $\rm  10 < \tau_V < 30$ & $\rm 2 \le \tau_V \le 6$ & 1.07 \\
   100 & 1000 & $\rm  1 < \tau_V < 20$ & $\rm 1 < \tau_V < 6$ & 1.07 \\
   100 & 50 & $\rm \tau_V > 30$ & $\rm 2 < \tau_V \le 6$ & 1.39 \\
   100 & 100 & $\rm 30 < \tau_V \le 50$ & $\rm 2 < \tau_V \le 6$ & 1.39 \\
   100 & 200 & & $\rm 1 < \tau_V < 6$ & 1.39 \\
   100 & 300 & $\rm 2 < \tau_V < 20$ & $\rm 1 < \tau_V \le 5$ & 1.39 \\
   100 & 500 & $\rm 1 < \tau_V \le 10$ & $\rm 1 \le \tau_V < 5$ & 1.39 \\
   100 & 1000 & $\rm 0.25 \le \tau_V < 8$ & $\rm 0.25 \le \tau_V \le 4$ & 1.39 \\
   100 & 50 & & $\rm 2 < \tau_V < 6$ & 1.85 \\
   100 & 100 & & $\rm 2 < \tau_V < 6$ & 1.85 \\
   100 & 200 & & $\rm 1 < \tau_V \le 4$ & 1.85 \\
   100 & 300 & $\rm 1 < \tau_V \le 5$ & $\rm 1 \le \tau_V < 4$ & 1.85 \\
   100 & 500 & $\rm 0.25 < \tau_V \le 3$ & $\rm 0.25 \le \tau_V < 3$ & 1.85 \\
   100 & 1000 & $\rm \tau_V \le 2$ & $\rm \tau_V < 2$ & 1.85 \\
   1000 & 50 & & $\rm 3 \le \tau_V \le 6$ & 1.07 \\
   1000 & 100 & & $\rm 3 \le \tau_V < 8$ & 1.07 \\
   1000 & 200 & & $\rm 2 < \tau_V < 8$ & 1.07 \\
   1000 & 300 & & $\rm 2 < \tau_V < 8$ & 1.07 \\
   1000 & 500 & & $\rm 2 < \tau_V < 8$ & 1.07 \\
   1000 & 1000 & $\tau_V \sim 30$ & $\rm 2 < \tau_V < 8$ & 1.07 \\
   1000 & 50 & & $\rm 2 < \tau_V < 4$ & 1.39 \\
   1000 & 100 & & $\rm 2 < \tau_V < 5$ & 1.39 \\
   1000 & 200 & & $\rm 2 < \tau_V \le 5$ & 1.39 \\
   1000 & 300 & & $\rm 2 < \tau_V < 6$ & 1.39 \\
   1000 & 500 & & $\rm 2 < \tau_V < 6$ & 1.39 \\
   1000 & 1000 & & $\rm 2 \le \tau_V < 6$ & 1.39 \\
   1000 & 300 & & $\rm 2 < \tau_V < 3$ & 1.85 \\
   1000 & 500 & & $\rm 2 \le \tau_V \le 3$ & 1.85 \\
   1000 & 1000 & & $\rm 1 < \tau_V < 4$ & 1.85 \\
  \hline
  \end{tabular}
\end{table}

A dusty stellar system originated in a 100 Myr burst evolves towards
redder $\rm I_c-K$ colours when it ages, for fixed dust distribution
and amount of dust.
It also evolves towards brighter K magnitudes for the first 100 Myr
and towards fainter K magnitudes during the following post-starburst phase.
Conversely, it becomes fainter in K and redder in $\rm I_c-K$
when $\tau_V$ increases, for a fixed age.
In general, younger starbursts with $t_{burst} = \rm 100~Myr$
must be more attenuated in order to match the properties of observed EROs,
be their dusty medium two-phase clumpy or homogeneous (Fig. 5).
In the second case, a lower amount of dust is needed in order to produce
a given reddening (cf. Fig. 3).

A DSG with a 1 Gyr burst evolves towards brighter K magnitudes
and redder $\rm I_c-K$ colours when it ages, for fixed dust distribution
and amount of dust.
Conversely, it becomes fainter in K and redder in $\rm I_c-K$
when $\tau_V$ increases, for a fixed age.
Interestingly, starbursts with $t_{burst} = \rm 1~Gyr$ do not match
observed ERGs if their dusty medium is two-phase clumpy
(in the way we assumed in Sect. 2.2.2) if not at $z \sim 1$
and for an age of 1 Gyr and $\tau_V \sim 30$ (Fig. 6).
They also hardly satisfy the ERO selection criterion at any age,
if not for large amounts of dust (i.e., for $\tau_V > 20$).
Conversely, DSGs with long bursts and a homogeneous dusty medium
fall in the region populated by observed ERGs for low/intermediate amounts
of dust and almost any age, at any redshift between 1 and 2.
For these models the range in dustiness or age allowed by observations
increases with decreasing $z$.
We conclude that the model DSGs/DPSGs at $1< z < 2$ span very well
the region of the $\rm I_c-K$ vs. K colour--magnitude plane
populated by observed ERGs.

At this point it is interesting to note what happens if a 10 times lower mass
(i.e., equal to $\rm 10^{10}~M_{\sun}$) of primordial gas is transformed
into stars at a constant rate on a timescale of 0.1 or 1 Gyr.
In this case, we find that DSGs with a short (long) burst
and $\rm SFR = 10^2~M_{\sun}~yr^{-1}$ ($\rm 10~M_{\sun}~yr^{-1}$)
and their dusty post-starburst phases result to be too faint to be detected by
present ground-based K-band surveys, whatever their redshift,
dust amount and distribution are.
Hence we conclude that these surveys are not sensitive to dusty formation
of low-mass galaxies at $1< z < 2$ (if any).

\section{The $\bf I_c-K$ vs. $\bf J-K$ colour--colour diagram}

Hereafter we will discuss only the $\rm I_c-K$ vs. $\rm J-K$
colour--colour diagram since similar conclusions hold
for the $\rm R_c-K$ vs. $\rm J-K$ colour--colour diagram.

The PM00 classification criterion for ERGs is:
\[ J-K = 0.36~(I_c - K) + 0.46. \]
ERGs with given observed $\rm I_c-K$ colour are classified as Es (DSGs)
at $1 < z < 2$ if they have a bluer (redder) observed $\rm J-K$ colour
than this.

\subsection{Dust effects on the colours of passively evolving, old, dusty ellipticals}

Fig. 7 shows the distribution in the $\rm I_c-K$ vs. $\rm J-K$
colour--colour plane of dust-free (filled symbols) and dusty (empty symbols) Es
at $1 < z < 2$.
The short-dashed and solid lines represent, respectively,
the ERO selection function (i.e., $\rm I_c-K \ge 4$)
and the PM00 classification criterion for ERGs.

All the model Es obey the PM00 classification criterion for ERGs, i.e.,
they fall on the left side of the delimiting line (solid line) in Fig. 7.
This is re-assuring for dust-free Es, since the evolutionary models
examined by PM00 adopt the same stellar tracks, while those adopted here
(Maraston 1998) rely on a different set of stellar models.

The new result is that models of dusty Es still fall on the left of
the PM00 separation line, for reasonable upper limits to the amount of dust
present in these stellar systems.

In general, the inclusion of dust attenuation moves the dust-free models
in Fig. 7 towards redder $\rm I_c-K$ and $\rm J-K$ colours,
almost parallel to the age-sequence track, for fixed metallicity,
at a given redshift, and closer to the PM00 separation line,
whatever redshift and age of the model elliptical galaxy are.
Thus dust attenution in Es may explain very red $\rm I_c-K$ and $\rm J-K$
colours without the need of invoking extremely old ages or higher redshifts
(or higher metallicities -- cf. PM00).
\begin{figure}
  \includegraphics[width=84mm]{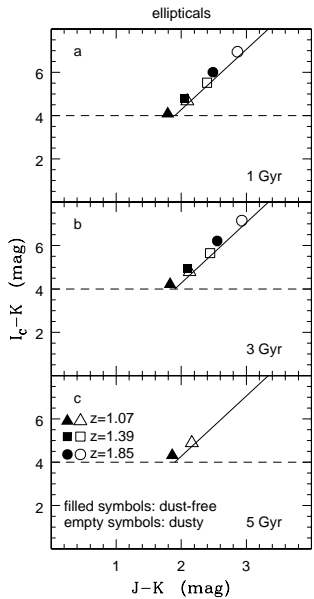}
  \caption{The distribution in the $\rm I_c-K$ vs. $\rm J-K$ colour--colour
plane of dust-free (filled symbols) and dusty (empty symbols) Es
at $1 < z < 2$. In each panel model Es of selected ages are reproduced
as triangles, squares and circles according to their redshifts.
The short-dashed and solid lines represent, respectively, the ERO
selection function and the PM00 classification criterion for ERGs.
All model Es obey the latter criterion.}
  \label{Fig. 7}
\end{figure}

\subsection{Dusty starburst/post-starburst galaxies}

Hereafter we refer to model DSGs with a 100 Myr burst and model DPSGs,
both with a two-phase clumpy distribution of the dust,
as {\it basic model DSGs} and {\it basic model DPSGs}, respectively,
given the results of GCW97.

\subsubsection{DSGs with a short burst and DPSGs}

Fig. 8 reproduces the distribution of basic model DSGs/DPSGs
with different age and dust amount in the $\rm I_c-K$ vs. $\rm J-K$
colour--colour plane.
It shows that DSGs with a 100 Myr burst at $1 < z < 2$
obey the PM00 classification criterion, whatever the structure of
their dusty medium and dust amount are.

Conversely, an intermediate age, dusty post-starburst galaxy
may be mistaken for either a passively evolving, old elliptical
or a young, dusty starburst, both at $1 < z < 2$, as a function of
its redshift, when classified only on the basis of optical/near-IR colours
such as $\rm I_c-K$ (or $\rm R_c-K$) and $\rm J-K$.

In particular, basic model DPSGs with ages between 0.2 and 1 Gyr
populate the region (upper left quadrant in Fig. 8)
where only dust-free Es were expected to fall, according to PM00
(see also Mannucci et al. 2002).
These ``intruder'' basic model DPSGs are at $1.3 < z < 2$
and span a large range in $\tau_V$ ($0.25 \le \tau_V \le 40$),
according to their age.
These values of $\tau_V$ are not unplausibly large, especially when
the model DPSGs have a homogeneous dust distribution (Fig. 9).
In this case, the reddening effect per given amount of dust is maximized
(Fig. 3), so that these ``intruders'' have $\tau_V \le 5$ (according to
their redshift).
We stress that the ``intruder'' DPSGs may be luminous enough
to be observed in present surveys at $\rm K < 20.5$,
be their dust distribution homogeneous or two-phase clumpy (cf. Tab. 1).
Hence the presence of a dusty post-starburst phase at $1.3 < z < 2$
is a source of concern for the neat result of PM00.

Finally, we note that the mix of DPSGs with ages between 0.2 and 1 Gyr
at $z \sim 1$ and DSGs (with ages between 0.05 and 0.1 Gyr) at $1 < z < 2$
in Fig. 8 is due to a conspiracy between redshift, shape of the intrinsic SED
and behavior of the attenuation function of these stellar systems.
The class of DPSGs is a suitable candidate for interpreting
the low dust emissions attributed to the bulk of dusty, bright EROs
on the basis of submm observations (cf. Introduction), as detailed in Sect. 5.
\begin{figure}
  \includegraphics[width=84mm]{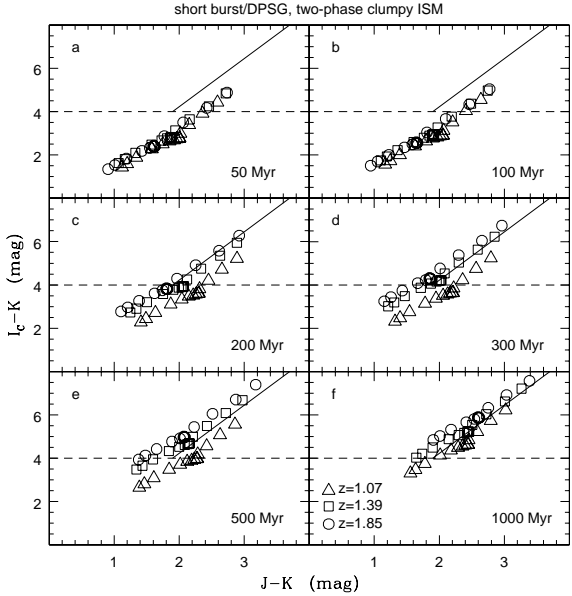}
  \caption{The distribution of the basic model DSGs/DPSGs in the $\rm I_c-K$
vs. $\rm J-K$ colour--colour plane, where lines are the same as in Fig. 7.
In each panel models of selected age are reproduced as triangles, squares
or circles according to their redshifts, with $\tau_V$ increasing
from 0.25 to 50 from the lower left to the upper right. Model DPSGs older
than 200 Myr and with a large range in $\tau_V$ not only populate
the region of DSGs (upper right quadrant) but they also ``intrude'' the region
(upper left quadrant) where only passively evolving, old, dust-free Es
were expected to fall (PM00).}
  \label{Fig. 8}
\end{figure}
\begin{figure}
  \includegraphics[width=84mm]{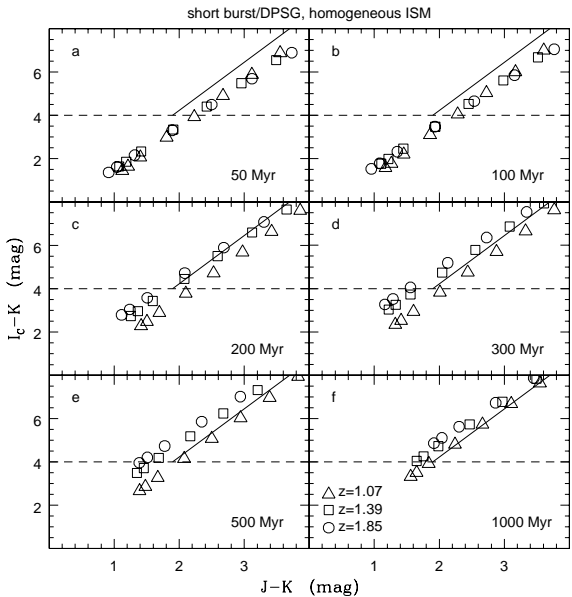}
  \caption{The distribution of DSGs with a short burst /DPSGs
with a homogeneous dusty medium in the same plane as in Fig. 8.
Each panel shows models of selected age at different $z$,
with $\tau_V$ increasing from 0.25 to 8 from the lower left to the upper right.
The ``intruder'' (see text) DPSGs with a homogeneous dusty medium
have stellar populations similar to those of basic model DPSGs in Fig. 8.}
  \label{Fig. 9}
\end{figure}
\begin{figure}
  \includegraphics[width=84mm]{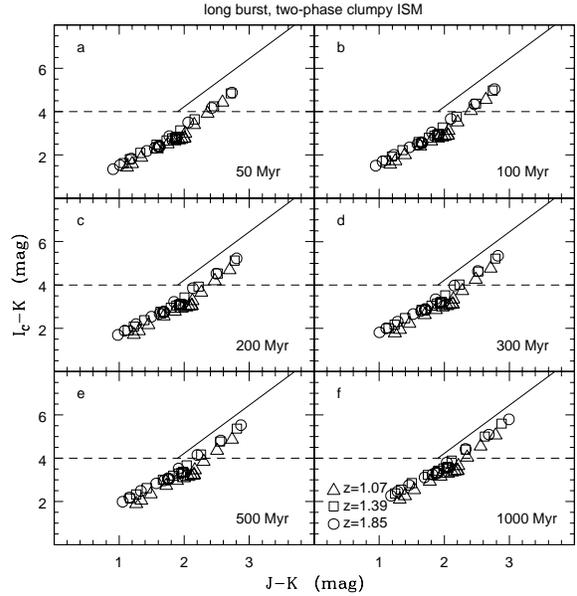}
  \caption{The distribution of model DSGs with a 1 Gyr burst
and a two-phase clumpy dust distribution in the $\rm I_c-K$
vs. $\rm J-K$ colour--colour plane as a function of age and $z$.
In each panel $\tau_V$ increases from 0.25 to 50 from the lower left
to the upper right. Lines and symbols are the same as in Fig. 8.
These DSGs meet the PM00 classification criterion for ERGs, even though
they are too faint to be selected as EROs (Fig. 6).}
  \label{Fig. 10}
\end{figure}
\begin{figure}
  \includegraphics[width=84mm]{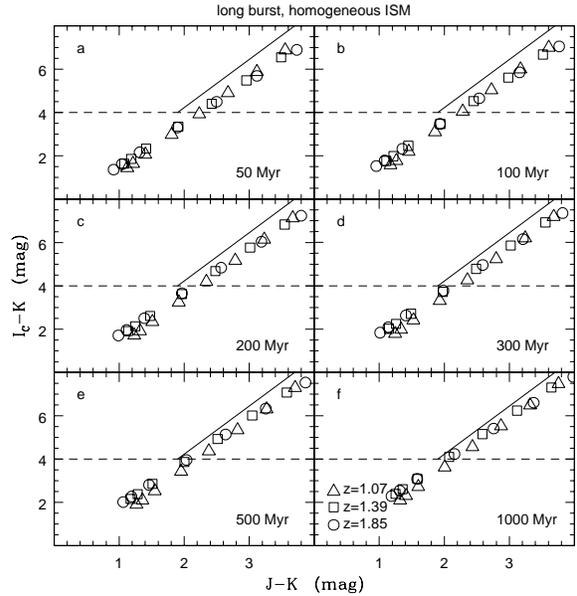}
  \caption{The same as in Fig. 10 but for a homogeneous dust distribution.
Each panel shows models of selected age at different $z$,
with $\tau_V$ increasing from 0.25 to 8 from the lower left to the upper right.
DSGs with a 1 Gyr burst and a homogeneous dusty medium meet
the PM00 classification criterion for ERGs. These DSGs may be bright enough
to be selected as EROs (cf. Tab. 1).}
  \label{Fig. 11}
\end{figure}

\subsubsection{DSGs with a long burst}

Figs. 10 and 11 show that model DSGs forming stars continuously in a long burst
satisfy the PM00 classification criterion, be their dust distribution
two-phase clumpy (but these models are too faint in K to be selected
as EROs -- see Fig. 6) or homogeneous, respectively.
PM00 have considered analogous starbursts observed during their continuous
star-formation activity.

Hence we confirm that there is a neat separation between passively evolving,
old stellar populations (in a dust-free environment)
and stellar populations formed in a continuous way in a dusty starburst,
both at $1 < z < 2$, (PM00) which holds whatever structure of the dusty medium
and dust amount of DSGs are.

We do not explicitly consider the case of early, dusty post-starburst phases
of a DSG with a long burst since the results obtained for basic model DPSGs
in Sect. 4.2.1 qualitatively apply to this case.
In particular, one may expect that this new class of DPSGs
presents ``intruders'' which have a lower dust amount than
``intruder'' basic model DPSGs observed at the same time after the end of
the burst and at the same redshift, owing to the existence of
the age/opacity degeneracy.
This is indeed what we find.
As an example, a 1.2 Gyr-old dusty post-starburst phase following a burst
with constant $\rm SFR = 10^2~M_{\sun}~yr^{-1}$ on a timescale of 1 Gyr
and observed at $z = 1.39$ has $\rm I_c-K = 4.44$ (5.10)
and $\rm J-K = 1.95$ (2.31), if its clumpy dusty medium has $\tau_V = 2$ (20).
This object falls well into the E domain (PM00) for $\tau_V = 2$
and, in general, is an ''intruder'' for low/intermediate values of $\tau_V$;
it falls into the DSG domain (PM00) for high values of $\tau_V$
(e.g. 20 or less so).
Conversely, a 0.3 Gyr-old basic model DPSG observed at $z = 1.39$
is an ``intruder'' when its clumpy dusty medium has $2 < \tau_V < 20$
(cf. Tab. 1).

\subsection{The case of sub-solar metallicities}

From our further modelling and from the analysis of PM00 (their Fig. 3c),
it is easy to realise that the previous conclusions extend to Es and DSGs/DPSGs
with metallicities different from solar.
For example, we find ``intruders'' among DPSGs with a two-phase clumpy
distribution of the dust and a tenth-solar metallicity (not shown).
In this case, the ``intruders'' are older than 300 Myr and have values of
$\tau_V$ between 0.50 and 30 (according to their $z$).

\subsection{The SED of an ``intruder'' DPSG}

Here we give an example of the non-negligible subsample of computed models
of Es and DPSGs which populate the same region
in the $\rm I_c-K$ vs. $\rm J-K$ colour--colour plane (cf. Figs. 7, 8 and 9).
Fig. 12 shows the normalized SED of an SSP (identified as a passively evolving
elliptical galaxy) with $\rm Z_{\sun}$ and age equal to 4 Gyr at $z = 1.39$
(thick solid line) and the normalized SED of a basic model DPSG
with $\rm Z_{\sun}$, age equal to 500 Myr and $\tau_V = 6$ at $z = 1.85$
(thin solid line).
Each SED is in the observed frame and is normalized to the luminosity
of the object at $\rm 2.17~\mu m$.
The spectral coverages of the $\rm I_c$, J and K filters are shown
as dashed areas.
The visual inspection of Fig. 12 shows that these two objects
have identical $\rm I_c-K$ and $\rm J-K$ colours.
Indeed $\rm I_c-K = 5.01$ and $\rm J-K = 2.11$ for the elliptical galaxy
and $\rm I_c-K = 4.97$ and $\rm J-K = 2.09$
for the DPSG\footnote{The elliptical galaxy has $\rm K = 20.73$
(for a stellar mass of $\rm 7.5 \times 10^{10}~M_{\sun}$),
while the DPSG has $\rm K = 21.04$. Hence these two models correspond to
objects beyond the limiting K-band magnitude of the present ERO samples.
N.B.: this is only a matter of normalization.}.

In theory, complementary optical photometry in U, B and V
may help distinguishing DPSGs at $1.3 < z < 2$ from Es at $1 < z < 2$,
which can not be distinguished by the PM00 classification criterion.
Model Es have e.g. values of $\rm V-R_c$ ranging from 2 to 3
when their $\rm V-K$ ranges from 7.5 to 12, the former range
being roughly 1 mag redder than that of model DPSGs
with the same $\rm V-K$ colour.
However, our models of the oldest Es and of the oldest, dustiest DPSGs,
both at $1 < z < 2$, predict apparent magnitudes fainter than 28 V-mag,
so that only the brightest ($\rm 21 < V < 26$) among these objects
(either non-extremely attenuated DPSGs at $1 < z < 2$
or very massive Es at $z \sim 1$), seem to be easily detectable,
given the current observational capabilities.

Hence we conclude and confirm that well-sampled SEDs are needed
for a robust classification of ERGs (e.g. Smith et al. 2001;
Smail et al. 2002; Miyazaki et al. 2002), as well as observations in X-rays,
radio and/or far-IR/submm (see Introduction).
Photometry in the far-IR/submm is suggested almost by default,
even though present observational campaigns with SCUBA (Andreani et al. 1999;
Mohan et al. 2002) show that the bulk of bright ($\rm K < 19.5$--20) ERGs
has total mid-IR--mm rest-frame luminosities lower than those of ULIRGs.
We discuss this in Sect. 5.

Finally, a way of distinguishing DPSGs from Es, both at high-$z$,
may be provided by the presence/absence of intermediate-age
(i.e., $10^8$--$10^9$ yr-old) stellar populations.
We will discuss this in a future paper.
\begin{figure}
  \includegraphics[width=84mm]{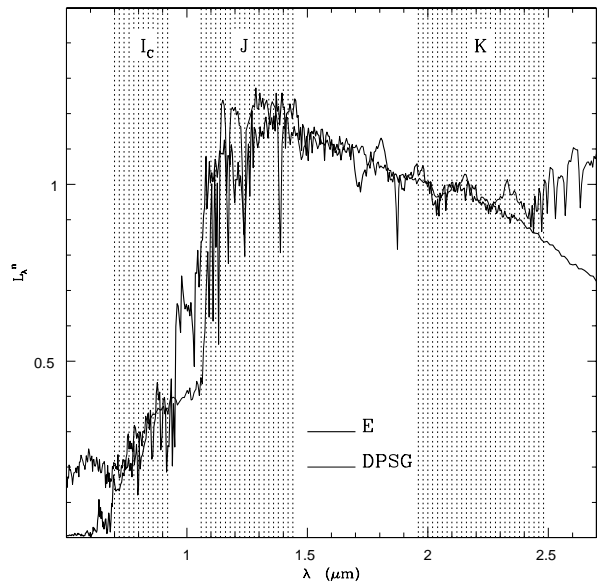}
  \caption{The SED of an SSP (i.e, an elliptical galaxy) with $\rm Z_{\sun}$
and age of 4 Gyr at $z = 1.39$ (thick solid line) and the SED
of a basic model DPSG with $\rm Z_{\sun}$, age of 500 Myr
and $\tau_V = 6$ at $z = 1.85$ (thin solid line). Each SED is in the observed
frame and is normalized to the luminosity at $\rm 2.17~\mu m$.
The spectral coverages of the $\rm I_c$, J and K filters are shown
as dashed areas. It is evident that these two objects have similar $\rm I_c-K$
and $\rm J-K$ colours.}
  \label{Fig. 12}
\end{figure}

\subsection{On the classification of ERGs}

The existence of dusty ERGs mixed with dust-free objects
having passively evolving, old stellar populations (i.e., Es)
and vice versa has been recently claimed by Miyazaki et al. (2002),
following the application of the photometric redshift technique
to multi-wavelength photometry obtained for 247 EROs
(see Fig. 5 in their paper).
Evidence in favour of the diversity of ERGs comes also from
the multi-wavelength analysis of 68 EROs by Smail et al. (2002)
via a similar photometric redshift technique.
Nevertheless, these authors conclude that the PM00 classification criterion 
is reasonably effective, at least in the $\rm R_c-K$ vs. $\rm J-K$
colour--colour plane (see Fig. 5 in Smail et al.).

Fig. 13 reproduces the distribution in the $\rm I_c-K$ vs. $\rm J-K$
colour--colour plane of 45 ERGs out of the ERO sample of the previous authors.
This subsample comprises ERGs brighter than $\rm K = 20.5$ ($\rm 5 \sigma$)
and detected at least in the $\rm I_c$ or J band.
ERGs classified by Smail et al. as evolved (i.e., with passively evolving,
old stellar populations in a dust-free environment), dusty (i.e., starburst)
or other (i.e., with SEDs not matching those of the previous two classes)
are reproduced as empty circles, filled circles and empty triangles,
respectively.
The visual inspection of Fig. 13 shows that:
\begin{itemize}
\item
half of the dusty objects fall within $\rm 1 \sigma$ from
the PM00 separation line (on both sides).
The existence of potential confusion near this separation line
was stressed by PM00.
\item
Two objects out of the 12 ERGs classified by Smail et al. as galaxies
with evolved SEDs fall within $\rm 1 \sigma$ from the PM00 separation line
(on both sides).
\item
Two more ERGs with evolved SEDs fall on the right side of
the PM00 separation line.
They could be evolved galaxies at $z > 2$ (see PM00).
\item
ERGs with SEDs more complex than evolved- or dusty-type fall everywhere.
\end{itemize}
For passively evolving, old ellipticals and dusty starbursts at $1 < z < 2$,
the PM00 classification criterion, confirmed by our models
(see Sect. 4.1, 4.2 and 4.3), is statistically effective,
as pointed out by Smail et al. (2002).
Indeed also the median colours of these two galaxy populations
are separate (see Table 2 in Smail et al.).

The more complex distribution in Fig. 13 (see also Miyazaki et al. 2002)
is only partly accounted for by our theoretical exploration.
In fact we find that model DPSGs may overlap with model Es,
independent of their dust distribution (Sect. 4.2.1)
and metallicity (Sect. 4.3).

Peculiar systems (e.g. AGN-host galaxies, mergers) or bulge$+$disk galaxies
(with different composite stellar populations, affected by dust
in different ways) at $1 < z < 2$ may also be part of the ERO population.
In fact the latter objects have been found by Smith et al. (2002),
Miyazaki et al. (2002) and Yan \& Thompson (2003).

The set of models discussed here does not account for these dusty galaxies.
However we note that a passively evolving, old, dust-free population
associated with the bulge component of a bulge$+$disk galaxy may present redder
colours than the same population associated with a dust-free elliptical galaxy,
owing to the attenuation from the dust in the disk.
At the same time, the disk stellar population may behave like
a continuously star-forming DSG with low/intermediate SFRs (cf. PM00).
Hence a bulge$+$disk galaxy at $1 < z < 2$ may be expected to fall
on both sides of the PM00 delimiting line.
We defer to a future paper the investigation of normal bulge$+$disk galaxies
at $z \ge 1$ as candidate ERGs.
\begin{figure}
  \includegraphics[width=84mm]{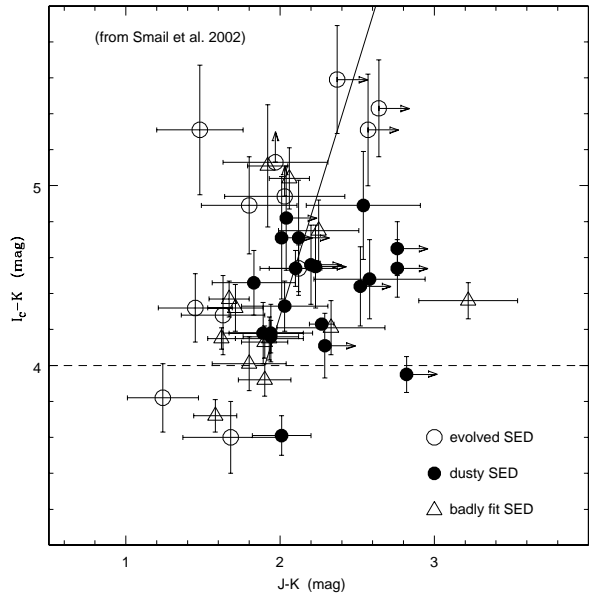}
  \caption{The distribution of 45 ERGs observed by Smail et al. (2002)
in the $\rm I_c-K$ vs. $\rm J-K$ colour--colour plane. ERGs classified by
Smail et al. as evolved, dusty or other are reproduced as empty circles,
filled circles and empty triangles, respectively. The short-dashed and solid
lines represent the ERO selection function and the PM00 classification
criterion for ERGs, respectively. Note that these objects were identified
as EROs by Smail et al. on the basis of their $\rm R_c-K$ colour.}
  \label{Fig. 13}
\end{figure}

\subsection[]{On the photometric definition of EROs}

Fig. 13 shows that 3 objects (with different classifications),
selected as EROs by Smail et al. (2002) on the basis of their $\rm R_c-K$
colour, do not match the $\rm I_c-K$ colour selection criterion of EROs
at more than $\rm 1 \sigma$, as already noted by these authors.
Smail et al. find that their field contains 68 EROs with
$\rm R_c-K \ge 5.3$ against 99 objects with $\rm I_c-K \ge 4$
and suggest that the latter is a less stringent definition of
an unusually red source.
In fact, 63 (93 per cent) of these 68 $\rm R_c-K$ colour-selected EROs
have $\rm I_c-K \ge 4$.

With our models, we can investigate the nature of these ``ambiguous'' objects.
We find that model Es at $1 < z < 2$, brighter than $\rm K = 20.5$,
obey both the $\rm R_c-K \ge 5.3$ and $\rm I_c-K \ge 4$ colour
selection criteria for EROs, whether dusty or not,
for both solar and half-solar metallicities (not shown).
Conversely, model DSGs/DPSGs at $1 < z < 2$, brighter than $\rm K = 20.5$,
selected via either $\rm I_c-K$ or $\rm R_c-K$, produce synthetic ERO samples
with different objects and sizes, for both solar
and a tenth-solar metallicities.
This is shown in Figs. 14 and 15 for $\rm Z_{\sun}$ DSGs with a short burst
and DPSGs, both with a two-phase clumpy dust distribution,
and for $\rm Z_{\sun}$ DSGs with a long burst
and a homogeneous dust distribution, respectively.

Interestingly, the $\rm R_c-K \ge 5.3$ colour selection criterion
systematically excludes more basic model DSGs/DPSGs at $1.3 < z < 2$
(mostly 300 Myr-old DPSGs with $0.50 \le \tau_V \le 10$,
but also DSGs with $\tau_V = 40$).
Conversely, the $\rm I_c-K \ge 4$ colour selection criterion
tends to exclude more DSGs/DPSGs at $z \sim 1$ (mostly 500 Myr-old DPSGs,
with $4 \le \tau_V \le 20$).
In case of model DSGs with a short burst and DPSGs with a homogeneous
dust distribution, the $\rm R_c-K \ge 5.3$ colour selection criterion
systematically excludes 300 Myr-old DPSGs with $\tau_V = 1$
and 500 Myr-old DPSGs with $\tau_V = 0.50$, both at $z > 1.8$.
Conversely, there is no evidence for a bias in $z$, caused by
the different colour selection criteria, for model DSGs with a long burst,
whatever the structure of their dusty medium is (not shown).
However, these DSGs are excluded by the $\rm R_c-K \ge 5.3$ colour selection
criterion if they have $\tau_V = 30$ (two-phase clumpy case)
or $2 \le \tau_V \le 3$ (homogeneous case).
Thus we conclude that selection via the $\rm R_c-K$ colour
bias the resultant ERO sample towards a larger ratio of Es
(at any $z$ between 1 and 2) to DSGs/DPSGs at $1.3 < z < 2$;
vice versa, selection via the $\rm I_c-K$ colour bias it towards
a larger ratio of Es (at any $z$ between 1 and 2) to DSGs/DPSGs at $z \sim 1$.

In addition, we find that ERO selection via the $\rm R_c-K$ colour tends
to exclude DSGs/DPSGs brighter than $\rm K = 20.5$ at $1.3 < z < 2$,
if they have relatively low amounts of dust
(i.e., with $0.50 \le \tau_V \le 3$),
whatever the structure of their dusty medium is.

Hence large discrepancies may be expected in the total number of selected EROs
and in the statistical properties of the different sub-populations of ERGs
present in a given ERO sample, whether selection is via $\rm R_c-K$
or $\rm I_c-K$.
These considerations help understanding the 33 per cent mismatch
in the number of EROs selected via either $\rm R_c-K$ or $\rm I_c-K$
found by Smail et al. (2002).

We note that Yan \& Thompson (2003) conclude that the Es-to-DSGs ratio
in a given ERO sample is larger, when selection is via the $\rm R_c-K$ colour
instead of the $\rm I_c-K$ colour, on the basis of the observed
optical/near-IR colours expected for model stellar populations
(Bruzual \& Charlot\footnote{see ftp://gemini.tuc.noao.edu/pub/charlot/bca5.})
approximating a passively evolving, old elliptical or simulating galaxies
of different ages with extended star formation activity, all at $z \sim 1$.
They do not consider dust attenuation, however.
Our results support and complement their conclusion.
\begin{figure}
  \includegraphics[width=84mm]{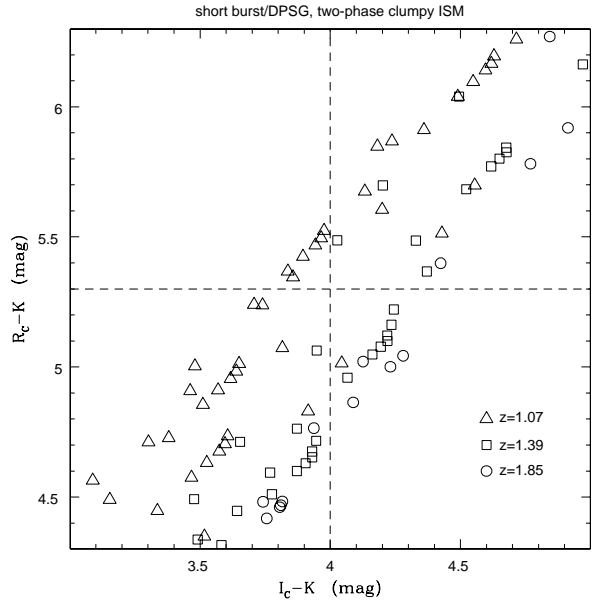}
  \caption{The distribution in the $\rm R_c-K$ vs. $\rm I_c-K$ colour--colour
plane of dusty stellar systems formed in a short burst, of $\rm Z_{\sun}$
and different ages, at $1 < z < 2$, brighter than $\rm K = 20.5$
and with a two-phase clumpy dust distribution. The two short-dashed lines
represent the $\rm R_c-K \ge 5.3$ and $\rm I_c-K \ge 4$ selection functions
of EROs. ERO selection via either colour criterion produces samples
with different objects and sizes. A selection effect in $z$ is present.}
  \label{Fig. 14}
\end{figure}
\begin{figure}
  \includegraphics[width=84mm]{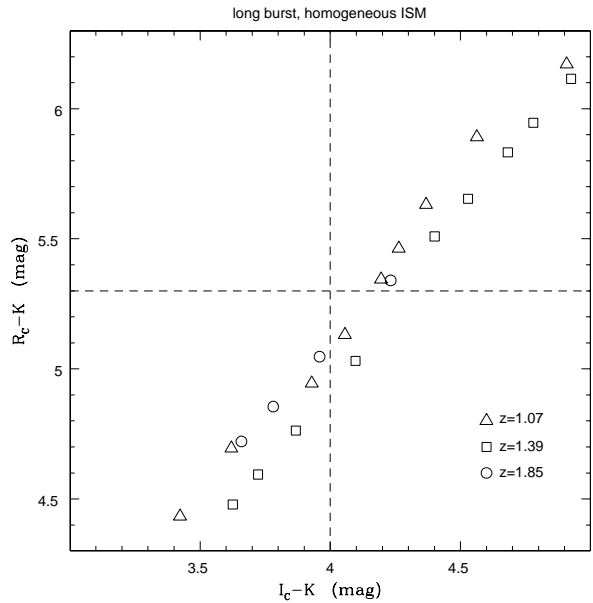}
  \caption{The distribution in the $\rm R_c-K$ vs. $\rm I_c-K$ colour--colour
plane of model DSG with a long burst, of $\rm Z_{\sun}$ and different ages,
at $1 < z < 2$, brighter than $\rm K = 20.5$ and with a homogeneous
dust distribution. Conclusions are the same as in Fig. 14, but for the absence
of a selection effect in $z$.}
  \label{Fig. 15}
\end{figure}

\section{Total mid-IR--(sub)mm emission of dusty starburst/post-starburst
galaxies at $\bf 1 < z < 2$}

Here we detail on the relation between amount of dust
and structure of the dusty medium, on one side, and total mid-IR--(sub)mm
emission from dust, on the other side, for model DSGs/DPSGs.
The radiative transfer models described in Sect. 2.2.2 do not model
re-emission from dust, and, thus, we can not produce and analyse SEDs
in the rest-frame mid-IR--(sub)mm for these stellar systems.
However, they provide a robust estimate of the total emission
of model DSGs/DPSGs at these wavelengths, as given by the amount of light
absorbed by internal dust in the wavelength range 0.1--$\rm 3~\mu m$
(in the rest-frame).
We refer to this estimate as $L_{dust}$.

We find that $L_{dust}$ has a dynamic range of more than 500
(from $\sim 10^{10}$ to $\rm \sim 5.6 \times 10^{12}~L_{\sun}$)
for DPSGs through DSGs, as a function of the dust parameter space.
In particular, $L_{dust}$ ranges
between 2 and $\rm 5.6 \times 10^{12}~L_{\sun}$ for DSGs with a short burst,
between 2 and $\rm 7.6 \times 10^{11}~L_{\sun}$ for DSGs with a long burst,
and between $10^{10}$ and $\rm 5.7 \times 10^{11}~L_{\sun}$ for DPSGs.
For fixed dust parameters, $L_{dust}$ increases (up to 50 per cent)
with increasing age for both DSGs, but decreases
(down to an order of magnitude) with increasing age for DPSGs.
Conversely, for fixed age, $L_{dust}$ increases
by a factor of about three when $\tau_V$ increases from 0.25 to 50
(from 0.25 to 8) for the two-phase clumpy (homogeneous) case,
for both DSGs and DPSGs.

The reason for the high dynamic range in $L_{dust}$ is twofold.
On one hand, a dusty medium with given dust type absorbs stellar light
at a given wavelength with an efficiency which strongly depends on
dust amount, structure of the dusty medium, and distribution of this medium
with respect to the stellar sources.
Hence all these aspects contribute to determining
the re-emission from dust grains in the mid-IR--(sub)mm wavelength domain.
This is investigated in details by Misselt et al. (2001)
for the same dust/stars geometry of a DSG/DPSG as that assumed in Sect. 2.2.2.
Here it is important to stress that model DSGs/DPSGs selected as ERGs
have very different dust properties (cf. Table 1)\footnote{In particular,
we note that structure of the dusty medium may be traded for dust amount
in producing a given value of $L_{dust}$. This is a caveat against
the simplistic use of dust emission as a proxy of dust amount in dusty ERGs.}.

On the other hand, dust absorbs and scatters more efficiently
UV photons than optical/near-IR ones.
Available optical/near-IR photons may replace available UV photons
as the major source of dust heating, if the optical depth
at optical/near-IR wavelengths of the dusty medium is large enough.
For a model DPSG, the intrinsic UV-to-optical luminosity ratio decreases
towards 0 when the model age approaches 1 Gyr (Fig. 1).
Thus ageing DPSGs will tend to have lower and lower
total mid-IR--(sub)mm luminosities, for fixed dust parameters.

For the dust parameter space explored here, only model DSGs
with $\rm SFR = 10^3~M_{\sun}~yr^{-1}$ over a timescale of 100 Myr
may be associated with $L_{dust} \ge \rm 10^{12}~L_{\sun}$
and, thus, be classified as ULIRGs.
Conversely, model DSGs with $\rm SFR = 10^2~M_{\sun}~yr^{-1}$
over a timescale of 1 Gyr typically have total mid-IR--(sub)mm luminosities
of a few to several $\rm \times 10^{11}~L_{\sun}$.
DPSGs may be associated with even lower values of dust emission
(down to $\rm \sim 10^{10}~L_{\sun}$), especially when their post-burst
stellar population approaches 1 Gyr.
Hence we conclude that the bulk of model DSGs/DPSGs does not belong to
the ULIRG class.
This is true also for model DSGs/DPSGs selected as EROs
with $\rm K < 20.5$ (cf. Table 1) or brighter (cf. Figs. 5 and 6).

Our conclusion is consistent with those of Andreani et al. (1999)
and Mohan et al. (2002), which are based on observational results
for a slightly brighter ERO sample (see Introduction).

\section{Discussion and conclusions}

In the previous sections we have modelled the spectral energy distributions
(SEDs) of simple stellar populations, intended to simulate
passively evolving, old ($\rm \ge 1~Gyr$) elliptical galaxies (Es),
and the SEDs of young ($\rm \le 0.1~Gyr$)/intermediate-age (0.1--1 Gyr),
dusty starburst galaxies (DSGs) and of intermediate-age, dusty post-starburst
galaxies (DPSGs), all at $1 < z < 2$, as candidate extremely red objects
(EROs).

In the early post-starburst phases, massive stars are absent
(owing to the absence of star formation) but intermediate-age stars
are present.
For these phases, we assume that at least part of the dusty medium
of the previous starburst phase survives during a maximal time-window
of 0.9 Gyr.

We have investigated observed optical/near-IR colours and magnitudes
of the previous models and, in particular, their distribution
in the $\rm I_c-K$ vs. $\rm J-K$ colour--colour plane.
In this colour--colour diagram, Pozzetti \& Mannucci (2000) claim that
passively evolving, old stellar populations in a dust-free environment
(i.e., elliptical galaxies) and stellar populations produced continuously
in a dusty starburst galaxy are reasonably well separated,
whatever the dust/stars configuration is.

This study complements the seminal analysis of Pozzetti \& Mannucci
for four main reasons.
\begin{itemize}
\item
First, we make the new case of {\it dusty post-starburst galaxies}
at $1 < z < 2$ as candidate EROs.
In particular, we discuss the early, dusty post-starburst phases of
stellar systems formed (mostly) at $1 < z < 2$, in a single burst of
finite duration ($t_{burst}$) of the order of $\rm 0.1~Gyr$
({\it short burst}).
The results obtained for these DPSGs may be easily extended to coeval,
early, dusty post-starburst phases of galaxy formation in a single burst
with e.g. $t_{burst} = \rm 1~Gyr$ ({\it long burst}),
when the existence of a degeneracy between age and opacity of the models
is taken into account.

We note that the DSGs with long or short bursts under investigation
sustain star formation rates (SFRs) of the order of $\rm 10^2~M_{\sun}~yr^{-1}$
or $\rm 10^3~M_{\sun}~yr^{-1}$, respectively.
SFRs $\rm \sim 10^3~M_{\sun}~yr^{-1}$ have been inferred
from observations of high-$z$ (sub)mm sources
(e.g., Smail, Ivison \& Blain 1997; Hughes et al. 1998; Barger et al. 1998;
Cimatti et al. 1998; Eales et al. 1999; Bertoldi et al. 2000; Lutz et al. 2001;
Smail et al. 2003).

The presence of a considerable population of starburst galaxies
with large SFRs over a short ($\rm \sim 100~Myr$) timescale at $1 < z < 2$
is qualitatively consistent with the interpretation of
the Lick absorption-line indices of nearby intermediate-mass
(i.e., with a velocity dispersion of $\rm \sim 200~km~s^{-1}$)
early-type galaxies by Thomas et al. (2002).
\item
Second, we explore the impact of dust attenuation (Witt et al. 1992)
on the optical/near-IR colours of dusty Es at $1 < z < 2$.
\item
Third, for all DSGs/DPSGs, we adopt a new dust/stars geometry
(Witt \& Gordon 2000) and explore the impact of dust distribution
(two-phase clumpy vs. homogeneous) and dust amount on the observed
optical/near-IR colours.
This geometry is consistent with observed properties of nearby starbursts
and Lyman Break Galaxies at $2 < z < 4$.
\item
Fourth, we calculate the total emission from dust
in the rest-frame mid-IR--(sub)mm for DSGs/DPSGs.
\end{itemize}
The distribution of model Es and DSGs/DPSGs at $1 < z < 2$
in the optical/near-IR colour--magnitude and colour--colour diagrams
well reproduces that of observed extremely red galaxies (ERGs) (Sect. 3 and 4).
Hereafter we summarise our results.
\begin{itemize}
\item
We find that DPSGs, with ages between 0.2 and 1 Gyr,
at $1.3 < z < 2$ may mix with Es at $1 < z < 2$
in the $\rm I_c-K$ vs. $\rm J-K$ colour--colour diagram,
for a large range in dust amount.
This holds whatever the structure (two-phase clumpy or homogeneous)
of their dusty medium is.
This result is a source of concern for the two-colour classification criterion
of Pozzetti \& Mannucci.
\item
Conversely, intermediate-age DPSGs at $z \sim 1$ may mix with DSGs
of age less than 1 Gyr at $1 < z < 2$ in the same colour--colour plot,
for a large range in dust amount and whatever the structure of
the dusty medium is.
This result and the previous one could explain in part the complexity of
the ERG classification recently emerged from the analysis of
statistical samples of EROs, where this classification was based on
multi-wavelength photometry (e.g. Smail et al. 2002; Miyazaki et al. 2002)
and not on optical/near-IR colour--colour plots.
\item
On the other hand, we confirm the result of Pozzetti \& Mannucci.
In addition, we find that this result holds
whether the dusty medium of a DSG is two-phase clumpy or homogeneous
and whether high-$z$ ellipticals are dust-free or as dusty as nearby ones.
\item
Model Es at $1 < z < 2$ populate a narrow locus
in the $\rm I_c-K$ vs. $\rm J-K$ colour--colour plane, whether dusty ot not.
This means that dust attenution may explain very red $\rm I_c-K$ and $\rm J-K$
colours for ellipticals at $1 < z < 2$ without the need of invoking
extremely old ages and/or higher metallicities and/or higher redshifts.
\item
The picture is the same when the distribution of the models
in the $\rm R_c-K$ vs. $\rm J-K$ colour--colour plane is investigated
(cf. Pozzetti \& Mannucci 2000).
Hence we conclude and confirm that well-sampled SEDs are needed
for a robust classification of ERGs (e.g. Smith et al. 2001;
Smail et al. 2002; Miyazaki et al. 2002), as well as observations in X-rays
(Alexander et al. 2002; Brusa et al. 2002; Smail et al. 2002;
but see Stevens et al. 2003), radio (Smail et al. 2002) and/or far-IR/submm
(Cimatti et al. 1998; Andreani et al. 1999; Mohan et al. 2002).
\item
We find that the total rest-frame mid-IR--(sub)mm luminosity
ranges from $\sim 10^{10}$ to $\rm \sim 5.5 \times 10^{12}~L_{\sun}$
for DPSGs through DSGs.
This high dynamic range is consistent with the large range
in total rest-frame mid-IR--(sub)mm luminosity estimated for ERGs
observed in the submm (Cimatti et al. 1998; Andreani et al. 1999;
Mohan et al. 2002).

This study offers two viable, not mutually exclusive interpretations
to the observational results and conclusions of the previous authors.
One interpretation is that re-emission by dust grains
in the mid-IR--(sub)mm wavelength domain depends on dust type and amount,
structure of the dusty medium and dust/stars configuration,
since the absorption efficiency of stellar light at given wavelength
depends on these parameters.
We find that the extremely red colours of DSGs/DPSGs at $1 < z < 2$
do not necessarily indicate large amounts of dust (Sect. 3),
as instead believed for e.g. SCUBA submm sources,
and may be reproduced under a very large range in structure of
the dusty medium.
Hence the submm fluxes of DSGs/DPSGs at $1 < z < 2$ are not expected
to be necessarily similar to those of SCUBA submm sources
at similar/higher $z$.

The other interpretation is the reduced contribution of UV photons
to dust heating in the model DPSGs, which seem to constitute
a non negligible, candidate ERO subpopulation.
Dust absorbs and scatters more efficiently
UV photons than optical/near-IR ones.
Available optical/near-IR photons may replace available UV photons
as the major source of dust heating if the optical depth
at optical/near-IR wavelengths of the dusty medium is large enough.
For a model DPSG, the intrinsic UV-to-optical luminosity ratio decreases
towards 0 when its age approaches 1 Gyr (Fig. 1).
Thus ageing DPSGs will tend to have lower and lower
total rest-frame mid-IR--(sub)mm luminosities, for fixed dust parameters.
\item
Furthermore, for the dust-parameter space considered here,
we find that only model DSGs with $\rm SFR = 10^3~M_{\sun}~yr^{-1}$
over a timescale of 100 Myr may be associated with a total mid-IR--(sub)mm
luminosity (in the rest frame) greater than $\rm 10^{12}~L_{\sun}$
and, thus, be classified as ULIRGs.
Conversely, model DSGs with $SFR = \rm 10^2~M_{\sun}~yr^{-1}$
over a timescale of 1 Gyr typically have total mid-IR--(sub)mm luminosities
of a few to several $\rm \times 10^{11}~L_{\sun}$.
DPSGs may be associated with even lower values of dust emission
(down to $\rm \sim 10^{10}~L_{\sun}$).
We conclude that the bulk of dusty, bright ($\rm K < 20.5$) ERGs
is not associated with ULIRGs, consistent with the conclusion of
Andreani et al. (1999) and Mohan et al. (2002), based on observations
of a slightly brighter sample.
\item
Typical ground-based surveys with $\rm K < 20.5$
(e.g., Drory et al. 2001; Cimatti et al. 2002b; Smail et al. 2002)
are biased against the following objects at $1 < z < 2$ (if any):
ellipticals (dusty or not) with stellar masses lower
than $\rm \sim 10^{11}~M_{\sun}$;
dusty starbursts with e.g. a constant $\rm SFR \le 10^2~M_{\sun}~yr^{-1}$
($\rm 10~M_{\sun}~yr^{-1}$) on a timescale of 0.1 (1) Gyr,
identified as events of low-mass galaxy formation at these redshifts;
early, dusty post-starburst phases of these dusty starbursts.
\item
Model Es (dusty or not) at $1 < z < 2$ are always selected as EROs
(if massive enough), whether selection is
via the $\rm I_c-K$ or $\rm R_c-K$ colour.
\item
Conversely, selection of observable (i.e., with $\rm K < 20.5$),
coeval DSGs/DPSGs as EROs is a complex function of the properties of
their dusty medium.
The explanation is that the extremely red colours
($\rm R_c-K \ge 5.3$ or $\rm I_c-K \ge 4$) of model DSGs/DPSGs at $1 < z < 2$
derive from radiative transfer through their dusty media.
\item
We find that selection via the $\rm R_c-K$ colour bias the resultant sample of
observable EROs towards a larger ratio of Es (at any $z$ between 1 and 2)
to DSGs/DPSGs at $1.3 < z < 2$; vice versa, selection via
the $\rm I_c-K$ colour bias it towards a larger ratio of Es
(at any $z$ between 1 and 2) to DSGs/DPSGs at $z \sim 1$.
This result confirms and complements that of Yan \& Thompson (2003).

Hence important discrepancies may be expected in the total number of
selected EROs and in the statistical properties of
the different sub-populations of ERGs present in a given ERO sample,
whether selection is via $\rm R_c-K$ or $\rm I_c-K$.
For example, DSGs/DPSGs at $1.3 < z < 2$ with the lowest amounts of dust
are not selected as EROs via the $\rm R_c-K \ge 5.3$ colour criterion.

These considerations help understanding the 33 per cent mismatch
in the number of EROs with $\rm K < 20.5$ selected via either $\rm R_c-K$
or $\rm I_c-K$ found by Smail et al. (2002).
\end{itemize}

\section*{Acknowledgments}

We acknowledge the anonimous referee, who contributed to the improvement of 
this paper in its final version with her/his stimulating, insightful comments
and suggestions.

\appendix

\section[]{The non-importance of the nebular continuum}

We reproduce the distribution in the $\rm I_c-K$ vs. $\rm J-K$
colour--colour plane of $\rm Z_{\sun}$ DSGs with a short burst and DPSGs
(Fig. A1) and DSGs with a long burst (Fig. A2), all of them
with a two-phase clumpy dust distribution, as a function of age and redshift,
when no nebular emission is computed.
Hence no attenuation of the gas emission, either in the line or
in the continuum, from dust is computed as well.
Models are represented as in Figs. 8 and 10, respectively,
where nebular emission was taken into account together
with its attenuation from dust as described in Sect. 2.2.2.

From the comparison of these four figures it emerges that
nebular emission and its attenuation do not play a role
in the selection of a DSG/DPSG at $1 < z < 2$ as an ERO.
They do not affect the presence of ``intruder'' DPSGs either.
Hence we conclude that the main results in Sect. 4.2 do not depend on
our choice for the attenuation of the nebular emission from dust.
\begin{figure}
  \includegraphics[width=84mm]{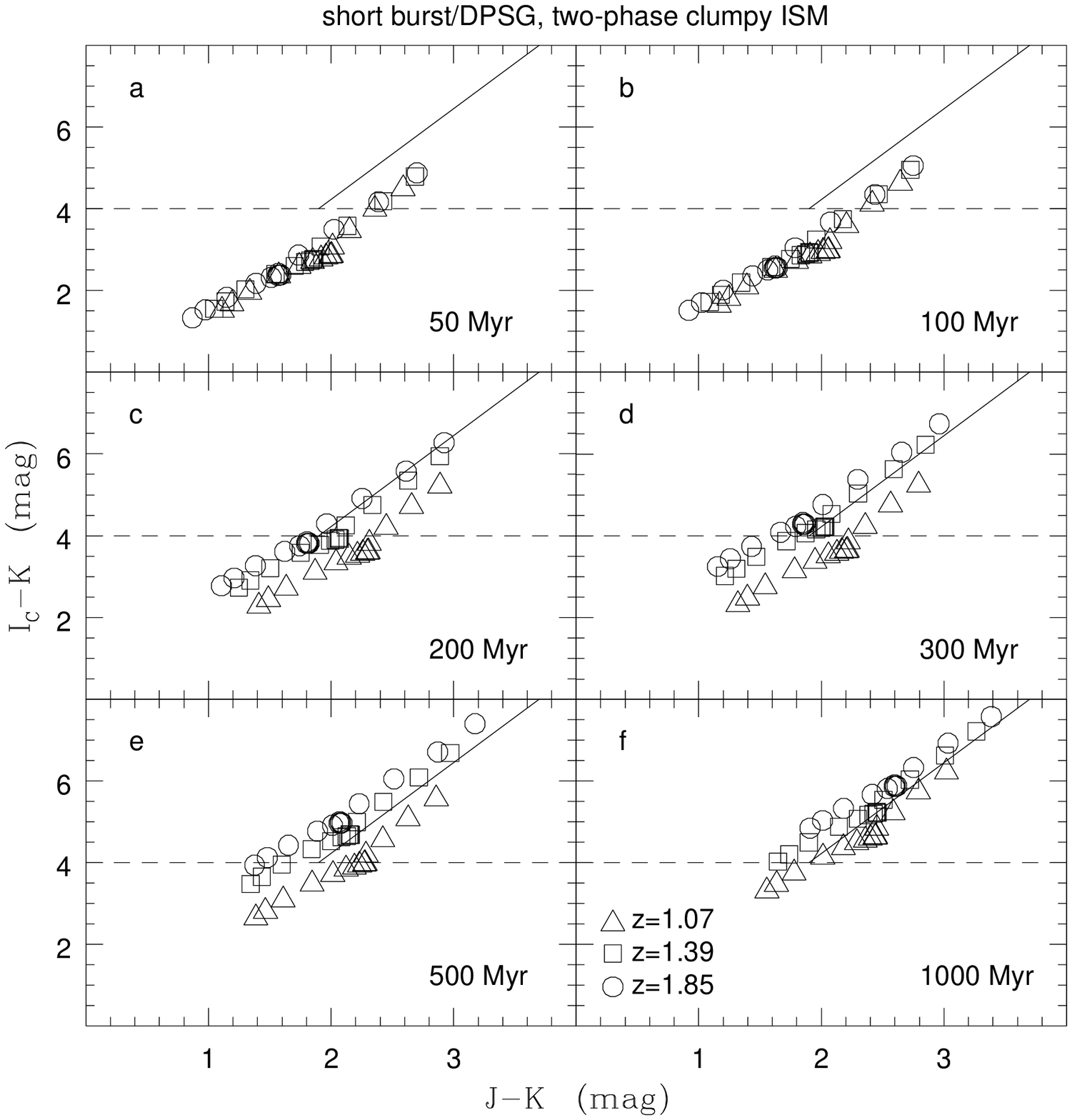}
  \caption{$\rm I_c-K$ vs. $\rm J-K$ for $\rm Z_{\sun}$ model DSGs
with a short burst and DPSGs, all of them with a two-phase clumpy dusty medium,
as a function of age and $z$, when no nebular emission is taken into account,
as well as its attenuation from dust. In each panel $\tau_V$ increases
from 0.25 to 50 from the lower left to the upper right. Lines and symbols
are the same as in Fig. 8. Conclusions are the same as those reached
from Fig. 8.}
  \label{Fig. A1}
\end{figure}
\begin{figure}
  \includegraphics[width=84mm]{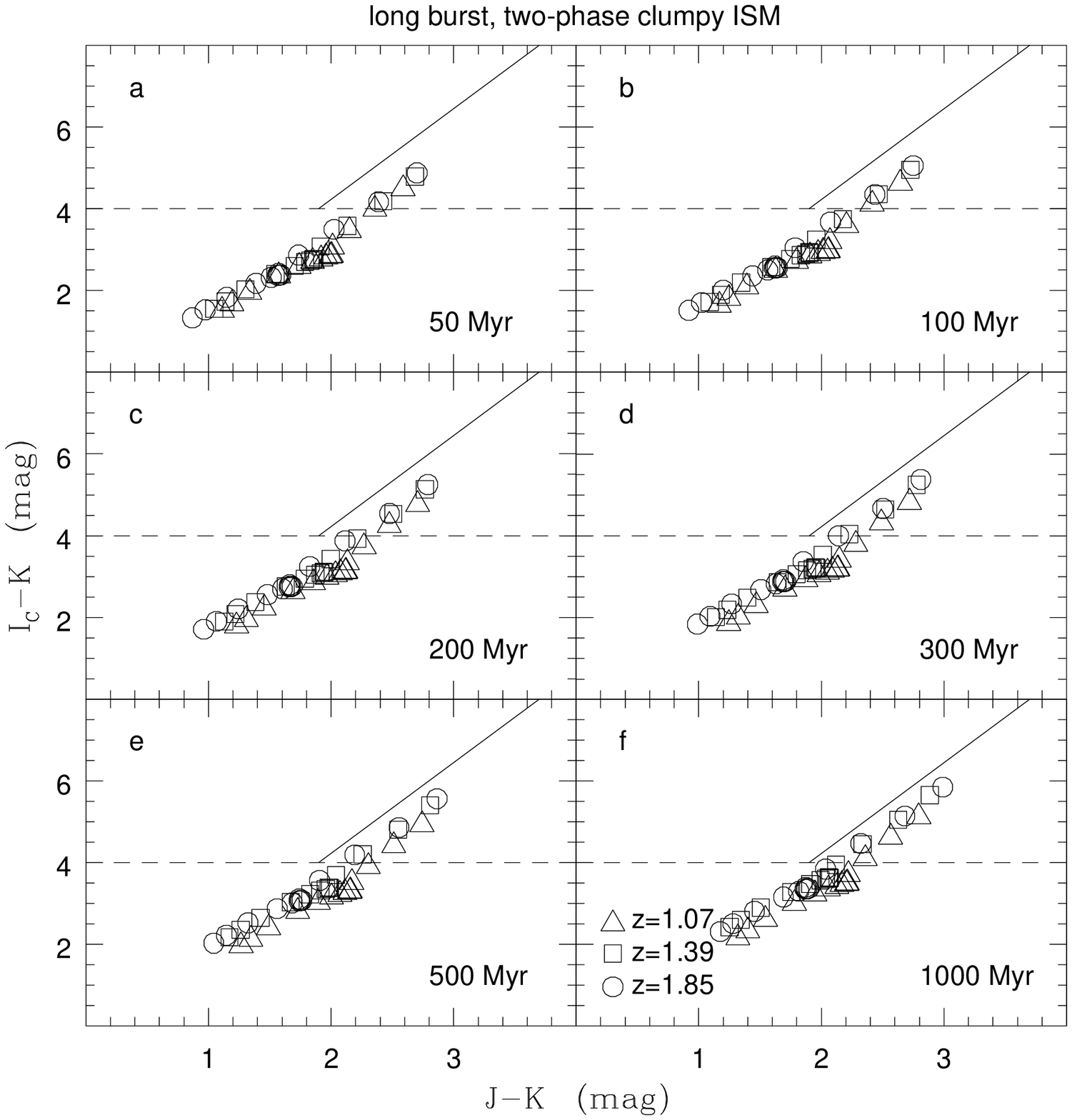}
  \caption{The same as in Fig. A1 but for DSGs with a long burst.
In each panel $\tau_V$ increases from 0.25 to 50 from the lower left
to the upper right. Conclusions are the same as those reached from Fig. 10.}
  \label{Fig. A2}
\end{figure}

\section[]{On the application of the Calzetti law to dusty ERGs}

Calzetti et al. (1994) have empirically constrained an extinction function
from the UV and optical spectra of 39 starburst and blue compact galaxies
at $z=0$.
This extinction curve has an overall UV/optical slope flatter than
the MW one and lacks the $\rm 2175~\AA$ dust absorption feature,
as confirmed by GCW97.
Further inclusion of IR broad-band data for 19 of these 39 galaxies
helped constrain the shape of the characteristic (i.e., {\it average})
attenuation function, but not fix its zero point (Calzetti 1997).
The absolute calibration of the attenuation function of DSG at $z=0$
was possible owing to the availability of far-IR photometry for a subsample
of only 5 objects (Calzetti et al. 2000).
In its final layout, the so-called ``Calzetti law'' has a fixed
wavelength dependence (as shown in Fig. 2) and is normalized
to the stellar-continuum reddening $E(B-V)_{star}$.
Since its discovery, the Calzetti law has been used to describe
dust attenuation in DSGs at high-$z$ (e.g., Cimatti et al. 2002a)
under the intrinsic assumption that dust properties and content,
as well as dust/stars distribution and structure of the dusty medium,
are similar in high-$z$ DSGs and in the nearby starbursts
of the Calzetti sample.

Thus we produce new models of $\rm Z_{\sun}$ DSGs with a short burst
and DPSGs (Fig. B1), and DSGs with a long burst (Fig. B2)
by adopting the Calzetti law instead of the simulated attenuation curves
described in Sect. 2.2.2.
Two vales of the stellar-continuum reddening (i.e., 0.5 and 0.8)
are considered, as in PM00 and consistent with Cimatti et al. (2002a).
Even in this case we find that model DPSGs partially ``intrude''
the region populated by passively evolving, old Es,
if they are ``moderately attenuated'' (i.e., $E(B-V)_{star} = 0.5$)
and at $1.3 < z < 2$ (Fig. B1).
Conversely, among the model DSGs with a long burst and attenuated
according to the Calzetti law, no ``intruders'' are found (Fig. B2).
Hence the main results of Sect. 4.2 are confirmed here.

There are no ``intruders'' with $\rm 4 \le I_c-K \le 5$ in Fig. B1,
as instead in Fig. 8, but this is easy to understand.
The original values of the Balmer line-emission reddening $E(B-V)$
found by Calzetti et al. (1994) for their sample range between 0 and 0.93,
so that $E(B-V)_{star}$ ranges between 0 and 0.41 for this sample,
if $E(B-V)_{star} = (0.44 \pm 0.03)~E(B-V)$ (Calzetti 1997).
Hence the observed values of $E(B-V)_{star}$ are lower
than those considered by PM00 and here.
The mean of these observed values is consistent with the value of
$E(B-V)_{star}$ predicted by the WG00 SHELL configuration with SMC-type dust
distributed in a two-phase clumpy medium and $\tau_V = 1.5$ (Fig. 3),
which reproduces rather well the Calzetti law (Fig. 2).
In Fig. 8 the ``intruders'' with $\rm 4 \le I_c-K \le 5$
have low/intermediate values of $\tau_V$ and, thus, $E(B-V)_{star} \le 0.3$,
consistent with the values obtained for the Calzetti sample.

Values of $E(B-V)_{star} \ge 0.5$ (e.g. in Cimatti et al. 2002a)
require either an enormous amount of dust, in the two-phase clumpy case,
or low amounts of dust and a homogenous dusty medium,
for the SHELL configuration (Fig. 3).
In both cases the wavelength dependence of the model attenuation function
is quite different from the Calzetti law (Fig. 2).
We conclude that the scaling of the functional form
giving the wavelength dependence of the Calzetti law with arbitrary values
of $E(B-V)_{star} \ge 0.5$, under the assumption that reddening is a measure
of dustiness, is not evinced from the observations and may be questionable
on a theoretical ground (cf. Sect. 2.2.2), as concluded also by WG00.
\begin{figure}
  \includegraphics[width=84mm]{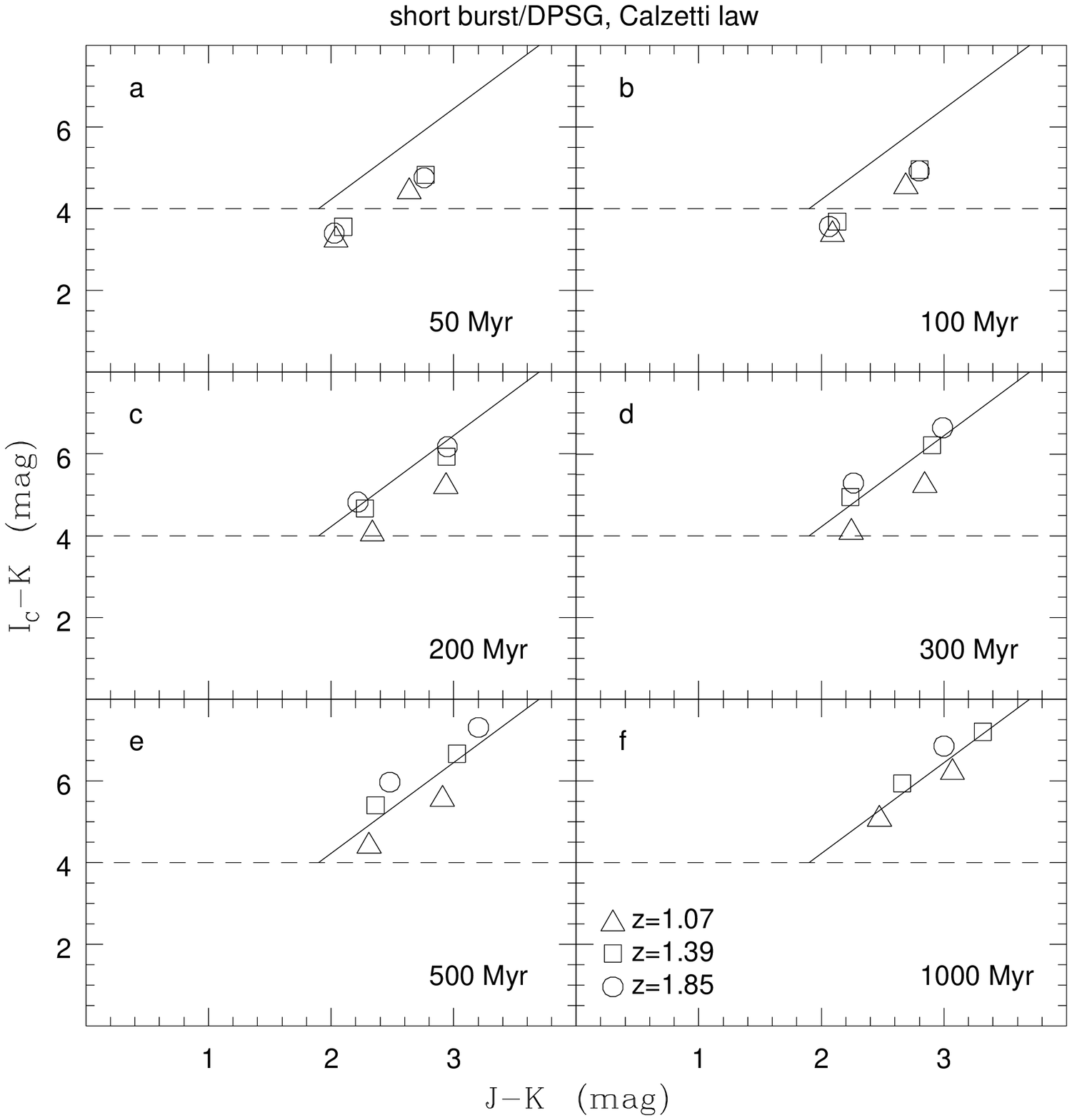}
  \caption{The distribution of $\rm Z_{\sun}$ DSGs with a short burst
and DPSGs in the $\rm I_c-K$ vs. $\rm J-K$ colour--colour plane
under the assumption that their attenuation function follows the Calzetti law.
Lines are the same as in Fig. 8. Models of selected ages are reproduced
with different symbols according to their redshift. DSGs/DPSGs with
a stellar reddening $E(B-V)_{star} = 0.5$ have bluer colours than those
with $E(B-V)_{star} = 0.8$. DPSGs attenuated with a Calzetti law ``intrude''
the region populated by passively evolving, old Es, consistent with the result
of Sect. 4.2.1.}
  \label{Fig. B1}
\end{figure}
\begin{figure}
  \includegraphics[width=84mm]{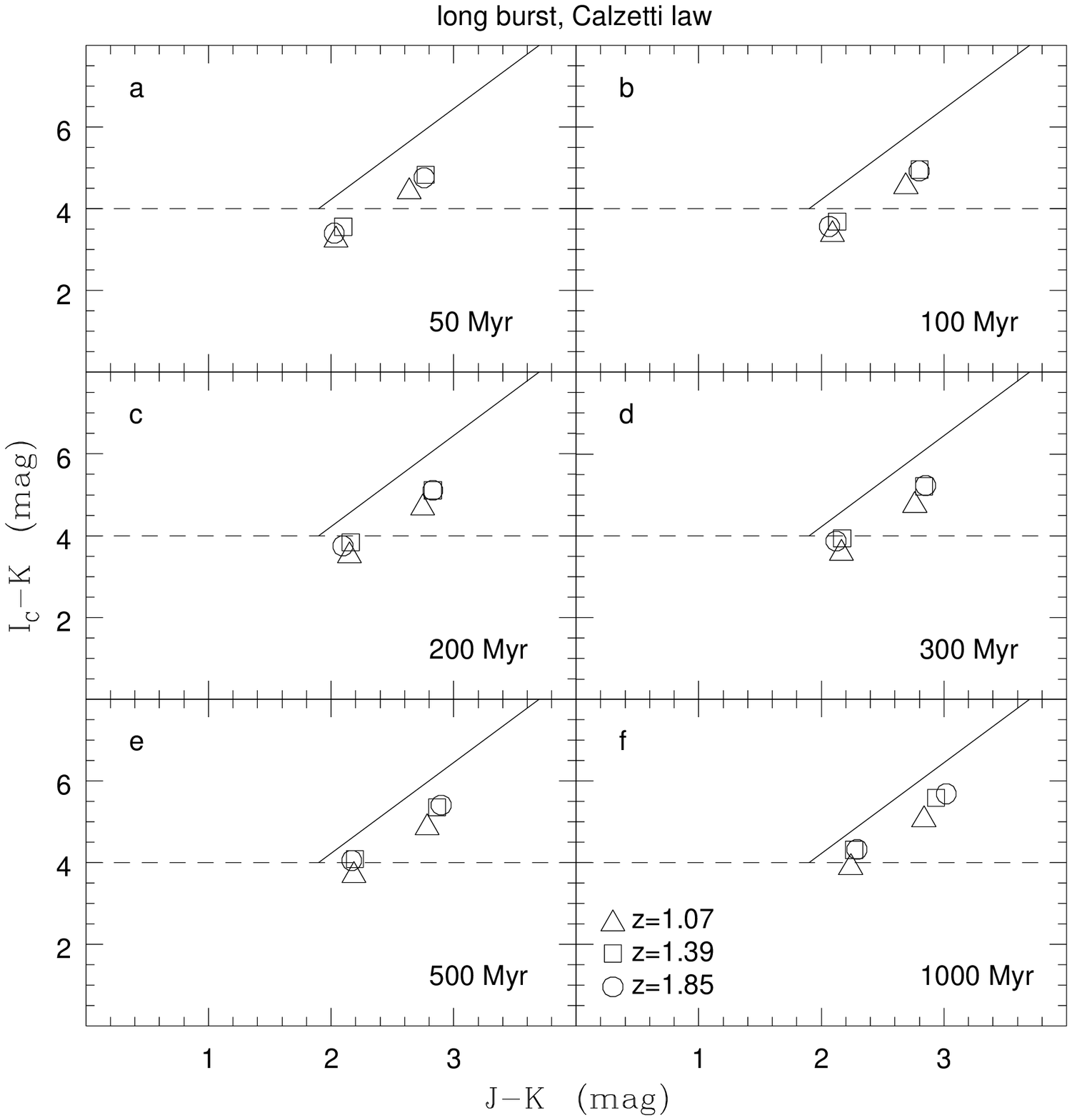}
  \caption{The same as in Fig. B1 but for $\rm Z_{\sun}$ DSGs
with a long burst. Models of selected ages are reproduced with
different symbols according to their redshift. In this case,
no ``intruders'' are found, consistent with the result of Sect. 4.2.2
and PM00.}
  \label{Fig. B2}
\end{figure}

\bsp

\label{lastpage}

\end{document}